\documentclass[fleqn,usenatbib]{mnras}

\usepackage{newtxtext,newtxmath}

\usepackage[T1]{fontenc}

\DeclareRobustCommand{\VAN}[3]{#2}
\let\VANthebibliography\thebibliography
\def\thebibliography{\DeclareRobustCommand{\VAN}[3]{##3}\VANthebibliography}

\usepackage{bm}
\usepackage{graphicx}	
\usepackage{amsmath}	
\usepackage{romannum}
\usepackage{float}
\DeclareUnicodeCharacter{2212}{-}

\newcommand{\Ep}{E_\nu\!\left(p\right)}
\newcommand{\fz}[1]{f^{(0)}\!\left(p#1\right)}

\title{Imprints of dark matter-massive neutrino interaction in upcoming post-reionization and galaxy surveys}

\author[A. Dey et al.]{Antara Dey,$^{1}$\thanks{E-mail: antaraaddey@gmail.com}
Arnab Paul$^{2}$\thanks{E-mail: arnabpaul9292@gmail.com} 
and Supratik Pal$^{1,\,3}$\thanks{E-mail: supratik@isical.ac.in}
\\
$^{1}$Physics and Applied Mathematics Unit, Indian Statistical Institute, Kolkata-700108, India\\
$^{2}$School of Physical Sciences, Indian Association for the Cultivation of Science, Kolkata-700032, India\\
$^{3}$Technology Innovation Hub on Data Science, Big Data Analytics and Data Curation,
	Indian Statistical Institute, Kolkata-700108, India\\}

\begin{document}
\label{firstpage}
\pagerange{\pageref{firstpage}--\pageref{lastpage}}
\maketitle

\begin{abstract}

We  explore possible signatures of the interaction between dark matter (DM) and massive neutrinos during the post-reionization epoch. Using both Fisher matrix forecast analysis and Markov Chain Monte-Carlo (MCMC) simulation,  we conduct a thorough investigation of the constraints and imprints of the scenario on the upcoming post-reionization and galaxy surveys. Our investigation focuses on two key parameters: the strength of the DM-massive neutrino interaction ($u$) and the total neutrino mass ($M_{\rm tot}$), on top of the usual 6 cosmological parameters. We utilize future 21-cm intensity mapping, galaxy clustering as well as cosmic shear observations in order to investigate the possible constraints 
of these parameters in the future observations: Square Kilometre Array (SKA1 and SKA2) and Euclid, 
taking both conservative and realistic approaches.  
All these missions show promise in constraining both the parameters $u$ and $M_{\rm tot}$  
by few orders compared to the current constraints from Planck18 (SKA2 performing the best among them). Although we do not find much improvement in $H_0$ and $\sigma_8$ tensions from our forecast analysis, SKA2 constrains them better in conservative approach.
We further perform a brief investigation of the prospects of some of the next generation Cosmic Microwave Background (CMB) missions 
in combinations with LSS experiments in improving the constraints. 
Our analysis reveals that both SKA2 and CMB-S4 + Euclid + SKA1 IM2 combination will put the strongest bounds on the model parameters.

\end{abstract}

\begin{keywords}
(cosmology:) dark matter - (cosmology:) cosmological parameters - (cosmology:) large-scale structure of Universe - cosmology: observations

\end{keywords}



\section{Introduction}
\label{Intro}

Cosmology in the next decade will largely be  driven by the world's largest radio telescope in the making Square Kilometre Array (SKA) \citep[]{Maartens:2015mra}, \citep[]{Santos:2015gra} and the upcoming  Large Scale Structure (LSS) mission Euclid \citep[]{EuclidTheoryWorkingGroup:2012gxx}, \citep[]{laureijs2011euclid} among others. They will survey a huge fraction of the sky and gather information from trillions of galaxies. The galaxy survey experiments like Euclid will achieve a tomography of the Universe over 13 billion years and will give precise knowledge of the Universe upto redshift 3. On the other hand, SKA will trace the neutral hydrogen in the Universe using 21-cm intensity mapping techniques and will provide enormous information of the yet-unavailable (barring some global signals) post-reionization and reionization epochs to the cosmic dawn  \citep[]{Cosmology-SWG:2015tjb} upto redshift 30. Apart from understanding the galaxy distribution and evolution as well as the physics of reionization, some of the major targets of these missions are to uncover the mysteries of dark matter, dark energy, neutrino mass and possible interactions between them, with an unprecedented accuracy. With the design and development of SKA going on in full swing and Euclid has just been launched, it is high time one explores the prospects of these missions in understanding these cosmic entities as clearly as possible.

Both Euclid and SKA will probe deep into non-linear regime which is governed  by non-linear physical processes like clustering of structures, 
 baryonic feedback \citep[]{vanDaalen:2011xb}, non-linear bias \citep[]{Jennings:2015lea} and non-linear misalignment of galaxies \citep[]{Takahashi:2012em}, \citep[]{Hilbert:2016ylf} along with other processes \citep[]{Casarini:2016ysv}. Like any other mission, it is imperative to define a  scale $k_{\rm max}$ in the non-linear regime, after which the noise of the instrument will supersede  the predicted signal. For Euclid the non-linear scale is $k_{\rm max} = 0.2 ~ \rm Mpc^{-1}$ \citep[]{EuclidTheoryWorkingGroup:2012gxx}, \citep[]{laureijs2011euclid}. Non-linear theoretical modelling of the processes are also available in the literature. However, from an instrumental point of view,  a conservative  approach \citep[]{Sprenger:2018tdb} to set the scale of $k_{\rm max}$ turns out to be more realistic, that we take up in this article.

While these surveys are our point of interest in the present article, so far as the model of the Universe is concerned, the $\Lambda$CDM model  is widely accepted as the {\textit {vanilla model}} among a plethora of models \citep[]{Cyr_Racine_2016}, \citep[]{Boddy_2018}, \citep[]{Barkana_2018}, \citep[]{Mangano_2006}, \citep[]{Wilkinson_2014}, \citep[]{Di_Valentino_2018}, \citep[]{Paul:2021ewd}, \citep[]{2023arXiv230501913A}, \citep[]{Blennow_2008} which is  mostly in tune  with  the latest observational data, barring some  tensions among different observations. In $\Lambda$CDM, the Universe is made up of $\sim 70 \%$ of cosmological constant $\Lambda$, $\sim 26 \%$ of cold dark matter (CDM) and $\sim 4 \%$ of baryonic matter. Here CDM is assumed to be non-relativistic and non-interacting with other species (except for gravitational interaction). A large number of observations starting from CMB \citep[]{refId0}, galaxies \citep[]{1970ApJ...159..379R}, clusters of galaxies \citep[]{Randall_2008}, to Large Scale Structures \citep[]{Tegmark_2004}, Type-Ia Supernovae data (SNIa) \citep[]{1998AJ....116.1009R}, \citep[]{1999ApJ...517..565P}, Lyman-alpha forest (Ly-$\alpha$) \citep[]{2002ApJ...581...20C}  all more or less conform with $\Lambda$CDM. However, from particle physics perspective, the exact nature of dark matter (DM) and its possible interaction with other relativistic/ non-relativistic species are yet unknown. However, if one relies on the particle DM, which is widely accepted in the community, it is interesting to allow possible non-gravitational interactions of DM, either with itself or with other fundamental particles, and investigate possible signatures in Cosmological/Astrophysical observations. Among the Beyond Standard Model (BSM) scenarios, one interesting possibility is Weakly Interacting Massive Particle (WIMP) where DM particles and standard model particles are in thermal equilibrium in the early Universe. This effect has its imprints on CMB anisotropy spectrum \citep[]{Slatyer_2009}, \citep[]{Chluba_2010} as DM particles dumps heat into standard model bath through annihilation. Other interesting possibilities are DM-dark radiation scattering \citep[]{Cyr_Racine_2016}, DM-baryons scattering \citep[]{Boddy_2018}, \citep[]{Barkana_2018}, DM-neutrino scattering \citep[]{Mangano_2006}, \citep[]{Wilkinson_2014}, \citep[]{Di_Valentino_2018}, \citep[]{Paul:2021ewd}, \citep[]{2023arXiv230501913A}, dark annihilation's into relativistic particles \citep[]{Blennow_2008}, DM-dark energy interaction \citep[]{Wang:2016lxa} etc.

In one of our previous work \citep[]{2023MNRAS.tmp.1769D}, we investigated the possible constraints on DM-massless neutrino scattering from the reionization era. We found that the reionization physics imposes more stringent constraints on the interaction parameter $u \leq 10^{-7}$ compared to the Planck18 data which puts $u \leq 1.003\times10^{-4}$ that was investigated by some of us earlier \citep[]{Paul:2021ewd} along with couple of other groups. However, as is well-known from neutrino oscillation experiments \citep[]{Cleveland_1998}, \citep[]{PhysRevLett.81.1562}, \citep[]{PhysRevLett.89.011301}, \citep[]{Pontecorvo:1967fh}, \citep[]{Gribov:1968kq}, \citep[]{lesgourgues_mangano_miele_pastor_2013} neutrinos do have finite mass, so including {\textit {massive}} neutrinos in the analysis turns out to be a more realistic scenario. The fluctuations of DM and neutrinos in the early phases of the Universe, when the perturbation theory holds, can be analyzed by means of the Boltzmann hierarchy equations. This particular aspect has been extensively investigated in previous studies \citep[]{Mosbech:2020ahp}. Including the neutrino mass in the analysis, the constraints on DM-massive neutrino interaction parameter  from Planck18 + SDSS BAO data has been found to be  $u \le 3.34\times10^{-4}$ \citep[]{Mosbech:2020ahp}.

However, the possible effects of post-reionization era on this interaction is yet to be explored in details.
The present work thus serves as a follow-up to our previous study \citep[]{2023MNRAS.tmp.1769D} as well as a natural extension of \citep[]{Mosbech:2020ahp}. In this article, we  investigate the interaction between DM and massive neutrinos thereby keeping the total neutrino mass as a free parameter in the analysis. We then investigate possible constraints on the scenario from post-reionization era using upcoming  missions SKA and LSS mission Euclid. To this end, we perform individual and joint Fisher forecast analysis to forecast on the errors as well as MCMC analysis to find the constraints on the model parameters. We further explore the prospects of upcoming CMB missions in constraining the interaction parameter. 
Our baseline model considers the lowest possible value for the total neutrino mass, accompanied by parameter values derived from the Planck18 data, thus forming our fiducial framework.

Throughout this work, we have adopted the fiducial values of the cosmological parameters as $w_\text{b}=0.02237,~w_\text{\rm nudm}=0.12010,~\ln[10^{10}A_\text{s}]=3.0447,~n_\text{s}=0.9659,~H_\text{0}=67.8,~\tau_\text{\rm reio}=0.0543,~ u=0, ~M_\text{\rm tot}=0.06$ eV where we have considered a minimum value for total neutrino mass and interaction parameter, other parameter values are consistent with Planck18 data \citep[]{refId0}.

\section{Effects of DM-massive neutrino interaction on linear perturbations
}\label{boltzman}

In the present framework, we treat DM (denoted by $\chi$) as a non-relativistic fluid, whereas for the neutrinos we consider a perturbed thermal distribution $f(\mathbf{x},\mathbf{p},\tau) = f_0 (p) \left[1+\Psi(\mathbf{x},\mathbf{p},\tau )\right]$, following the Boltzmann equation, $$\frac{d}{dt} f \left(\mathbf{x},\mathbf{p},t\right) = C\left[f \left(\mathbf{x},\mathbf{p},t\right)\right],$$ where $C\left[f \left(\mathbf{x},\mathbf{p},t\right)\right]$ is the collision term. For our present scenario, where there is a scattering process involved between the neutrinos and DM, the cosmological perturbation equations upto linear order and in Newtonian gauge  look like \citep[]{Mosbech:2020ahp},

\begin{subequations}
\begin{align}
    \dot{\delta}_{\chi} &= -\theta_\chi + 3\dot{\phi}~,\\
    \dot{\theta}_\chi &= -\frac{\dot{a}}{a}\theta_\chi + k^2 \psi~+ K_\chi \dot{\mu}_\chi \left(\theta_\nu - \theta_\chi\right)~,\\
    \frac{\partial \Psi_0}{\partial \tau} &=-\frac{pk}{\Ep}\Psi_1 - \dot{\phi} \frac{d\ln\fz{}}{d\ln p}~,\\
    \frac{\partial \Psi_1}{\partial \tau} &=\frac{1}{3}\frac{pk}{\Ep} \left(\Psi_0-2\Psi_2\right) - \frac{\Ep k}{3p}\psi\frac{d\ln\fz{}}{d\ln p} \nonumber \\ 
    &- C_\chi\frac{v_\chi E_\nu \! \left(p\right)}{3\fz{}} \frac{d\fz{}}{d p} - C_\chi\Psi_1~,\\
    \frac{\partial\Psi_2}{\partial\tau} &=
    \frac{1}{5}\frac{pk}{\Ep} 
    \left(2\Psi_{1}-3\Psi_{3}\right)- \frac{9}{10}C_\chi\Psi_2~, \\
    \frac{\partial\Psi_l}{\partial\tau} &=
    \frac{1}{2l+1}\frac{pk}{\Ep} 
    \left(l\Psi_{l-1}-(l+1)\Psi_{l+1}\right)-C_\chi\Psi_l, \quad l\geq3~.
\end{align}
\end{subequations}
where $K_\chi \equiv \frac{\rho_\nu + P_\nu}{\rho_\chi} = \frac{(1+w_\nu)\rho_\nu}{\rho_\chi}~$, $\dot{\mu}_\chi=a \sigma_0 c n_\chi$, $	C_\chi = a \, u \, \frac{\sigma_{\text{Th}} \rho_\chi}{100\, \text{GeV}}\frac{p^2}{E_\nu^2}~$ with 
 $u = \frac{\sigma_0}{\sigma_{\text{Th}}}\left(\frac{m_\chi}{100\, \text{GeV}}\right)^{-1}$. $\sigma_0$ is the scattering cross-section between DM and neutrinos. As per usual convention, $\phi$ and $\psi$ are the scalar metric perturbations, $\delta_\chi$ and $\theta_\chi$ are the over-density and velocity divergence of the DM species, $\Psi_l$ corresponds to the different modes of the Legendre expansion of $\Psi$, i.e. $\Psi\!\left(\mathbf{k},\hat{\mathbf{n}},p,\tau\right) = \sum_{l=0}^{\infty}\left(-i\right)^l \left(2l+1\right) \Psi_l \! \left(\mathbf{k},p,\tau\right) P_l \! \left(\hat{\mathbf{k}}\cdot\hat{\mathbf{n}}\right).$

In general, the scattering cross-section can be velocity-dependent. However, in the present analysis we will  consider $\sigma_0$ to be constant (s-wave scattering). 
 This dimensionless quantity $u$, quantifying the scattering strength, along with the sum of neutrino masses $M_{\rm tot}$ (we assume degenerate mass hierarchy in this work), are the only non-standard parameters on top of the 6-parameter vanilla $\rm \Lambda$CDM model. We use a  modified version \citep[]{Mosbech:2020ahp} of the publicly available code \texttt{CLASS} \citep[]{Blas_2011} to solve the perturbation equations.

\section{Observables and Mock Likelihoods in Future 21-cm and Galaxy Surveys}
For our analysis, we have taken into account three distinct types of future experiments in order to perform Fisher forecast followed by MCMC analysis on interacting DM-massive neutrino scenario. The individual experiments are as follows:
\begin{itemize}
\item 21-cm intensity mapping observations 
\item galaxy clustering observations 
\item cosmic shear observations. 
\end{itemize}
Below we explain each experiment considered in our analysis.

\subsection{Upcoming 21-cm Observations in Post-reionization Epoch}

After the reionization epoch, the Universe is almost completely ionized at a redshift $z<5$ with a neutral hydrogen fraction $ x_{\rm H1}< 10^{-5}$ \citep[]{Mitra_2015}, \citep[]{Kulkarni:2018erh} remaining within the galaxies. The sought after observable quantity in 21-cm cosmology is the differential brightness temperature $\Delta T_{\rm b}$, mapping the intensity of 21-cm line from the  leftover neutral hydrogen. The differential brightness temperature is defined as the difference between the spin temperature ($T_{\rm s}$) and the CMB temperature ($T_{\rm \nu}$). As our Universe is expanding, the line due to hyperfine transition of the neutral hydrogen with frequency 1420.4057 MHz emitted from a redshift $z$, gives rise to the observed $\Delta T_{\rm b}$ as \citep[]{Furlanetto:2006jb}, \citep[]{Pritchard:2011xb} 
\begin{equation}
    \Delta T_{\rm b}= \dfrac{T_{\rm s}(z)-T_{\rm \nu}(z)}{1+z}.
\end{equation}
For the present analysis that focuses on post-reionization epoch, we shall take into account only the low-redshift ($z \leq 3$) signals of the differential brightness temperature, coming from the neutral hydrogen within the galaxies. The mean brightness temperature can be approximated as \citep[]{Furlanetto:2006jb}, \citep[]{Pritchard:2011xb},

\begin{equation}
    \overline{\Delta T_{\rm b}}\approx 189 \left[\dfrac{H_{\rm 0}(1+z)^2}{H(z)}\right] \Omega_{\rm HI(z)}\textit{h}  ~\rm mK,
\end{equation}
Where $H_{\rm0}$ is the Hubble constant today $H_{\rm0}=h \times 100~ \rm km~s^{\rm -1} Mpc^{\rm -1}$ and $\Omega_{\rm HI}= \rho_{\rm HI}(z)/\rho_{\rm c}$ is the neutral hydrogen density parameter. 

The deviation of $\Delta T_{\rm b}$ from this mean $\overline{\Delta T_{\rm b}}$ can then be related to the matter density fluctuations $\delta_{\rm m}$ modulo an additional bias term $b_{\rm HI}$, which connects the  neutral hydrogen to the dark matter density contrast,
\begin{equation}
    \Delta T_{\rm b} - \overline{\Delta T_{\rm b}} = \overline{\Delta T_{\rm b}} \delta_{\rm HI} = \overline{\Delta T_{\rm b}} b_{\rm HI}\delta_{\rm m}.
\end{equation}

The power spectrum of this quantity is then related to the matter power spectrum via the relation,

\begin{equation}
    P_{\rm 21} = b_{\rm 21}^{2} P_{\rm m} = (\overline{\Delta T_{\rm b}} b_{\rm HI})^{2} P_{\rm m}.
\end{equation}

However, as we are mapping the neutral hydrogen in the galaxies, this theoretical power spectrum must be modified by redshift effects due to movement of galaxies, effects due to limited resolution and the Alcock-Paczinsky effect, described by $ f_{\rm RSD}(\hat{k},\hat{\mu},z)$, $f_{\rm res}(k,\mu,z)$ and $f_{\rm AP}(z)$ respectively, to match the observed 
 21-cm intensity power spectrum $P_{\rm 21}(k,\mu,z)$ \citep[]{Sprenger:2018tdb},
\begin{equation}
    P_{\rm 21}(k,\mu,z)= f_{\rm AP}(z) \times f_{\rm res}(k,\mu,z) \times f_{\rm RSD}(\hat{k},\hat{\rm \mu},z) \times b_{\rm 21}^{2}(z) \times P_{\rm m}(\hat{k},z)
\end{equation}
Here $P_{\rm m}(\hat{k},z)$ is the matter power spectrum. In this above formula we have incorporated the flat-sky approximation which gives distinctive definition of the line of sight distance vector $\Vec{r}$ and Fourier modes. Along the line of sight observer's fixed point violates the isotropic nature but the symmetry along perpendicular to the line of sight direction is preserved. Here these are the following relations to the coordinates, $k = |\Vec{k}|$, $\mu = \dfrac{\Vec{k}.\Vec{r}}{\rm k ~\rm r}$ and the parallel part of the mode is $k_{\rm \parallel}=\mu ~\rm k$ and perpendicular part is $k_{\rm \perp}=k\sqrt{1-\mu^{2}}$. Details about these various factors will be discussed in the next section. The major limitation of the 21-cm intensity mapping surveys are the interference of the signal with the telescope noise and foreground. If the foregrounds are sufficiently smooth in frequency, they can be completely removed from the HI signal. Assuming this, the observed power spectrum will only have contribution from the telescope noise $P_{\rm N}(z)$ \citep[]{Sprenger:2018tdb},
\begin{equation}
    P_{\rm 21,obs}(k,\mu,z) = P_{\rm 21}(k,\mu,z) + P_{\rm N}(z)
\end{equation}
Details about $P_{\rm N}(z)$ for a specific observation will be described in the next subsection. For a future observation, given the device specifications and assumptions about the fiducial values of the cosmological model, the mock data is generated using this expression.

\subsubsection{SKA}
SKA (Square Kilometer Array) is the largest radio-telescope in the world being built in Australia (low-frequency) and South Africa (mid-frequency). SKA will provide information about the neutral hydrogen in the Universe using 21-cm intensity mapping survey. The noise power spectrum for a survey in a single dish mode is given by \citep[]{Villaescusa-Navarro:2016kbz}, \citep[]{2021MNRAS.505.3492S}, 
\begin{equation}
    P_{\rm N}^{2} = T_{\rm sys}^{2} \dfrac{4\pi f_{\rm sky} r^{2}(z)(1+z)^{2}}{2H(z)t_{\rm tot}\nu_{0}N_{\rm dish}},
\end{equation}
where $T_{\rm sys}$ is the system temperature, $t_{\rm tot}$ is the total observation time and $N_{\rm dish}$ is the number of dishes. In the present work, we adopt $t_{\rm tot}$=1000 hours and $N_{\rm dish}$=200 as in \citep[]{Villaescusa-Navarro:2016kbz}. Foregrounds are projected to have much greater magnitude than the 21cm signal. Nevertheless, foregrounds are expected to exhibit spectral smoothness that enables their successful removal. In this particular study, we have incorporated a reduced sky fraction of $f_{\rm sky}=0.58$ and have accounted for a narrower frequency range \citep[]{2015MNRAS.447..400A} to address foreground removal. The instrumental specifications relevant for the intensity mapping are listed in Table \ref{skaIMspecs}.

\begin{table*}
\centering
\begin{tabular}{c|c|c|c|c|c|c}
\hline
    Parameter & $\nu_{\rm min}$($\rm MHz$) & $\nu_{\rm \max}$($\rm MHz$) & $z_{\rm min}$ & $z_{\rm max}$ & $\delta_{\rm \nu}$($\rm kHz$) & $T_{\rm inst}$($\rm K$)\\
\hline    
    SKA1 Band 1 & ~400 (350) & ~1000 (1050) & 0.45 & 2.65 & 10.9 & 23 \\
    SKA1 Band 2 & ~1000 (950) & 1421 (1760) & 0.05 & 0.45 & 12.7 & 15.5
\\
\hline

\end{tabular}  
\caption{SKA Intensity Mapping specifications 
\citep[]{Olivari:2017bfv}.}
		\label{skaIMspecs}
 
\end{table*}

\subsection{Upcoming Galaxy Clustering Observations}

The galaxy surveys provide us with the spatial distribution of galaxies. Apart from the fact that this distribution  is a biased tracer of the underlying DM distribution of the Universe, similar to the 21-cm power spectrum, there are various astrophysical and astronomical factors that have to be taken into account to estimate galaxy power spectrum $P_{\rm g}$ from DM power spectrum $P_{\rm m}$ \citep[]{Sprenger:2018tdb},
\begin{equation}
    P_{\rm g}(k,\mu,z)= f_{\rm AP}(z) \times f_{\rm res}(k,\mu,z) \times f_{\rm RSD}(\hat{k},\hat{\mu},z) \times b^{2}(z) \times P_{\rm m}(\hat{k},z).
    \label{eq:1}
\end{equation}
We have considered flat sky approximation \citep[]{Lemos:2017arq}, \citep[]{Asgari:2016txw} for generating galaxy power spectrum. As discussed in the previous section, the factor $f_{\rm AP}$
is the contribution from the Alcock-Paczinsky effect \citep[]{1979Natur.281..358A}, \citep[]{2003ApJ...598..720S}, which takes into account the difference between the true cosmology and assumed or fiducial cosmology, $f_{\rm AP}(z)=D_{\rm A}^{2} \rm H_{\rm t} / D_{\rm A,t}^{2} \rm H$, where $D_{\rm A}$ and $H$ are the angular diameter distance and Hubble parameter and 't' denotes the true cosmology. As mentioned previously, the second and the third terms are contributions from the resolution effect due to limited resolution of the instruments 
and redshift distortion respectively.

\subsubsection{Euclid} Euclid is a European Space Agency mission, from which we expect to get further insights about dark matter distribution and dark energy of the Universe. Using spectroscopic survey, Euclid will map about $10^{7}$ galaxies within a redshift range of $0.7-2.0$. We have considered the error on spectroscopic measurement survey to be $\sigma_{\rm z} = 0.001(1+z)$ \citep[]{EuclidTheoryWorkingGroup:2012gxx}, \citep[]{Audren:2012vy} and we have neglected the error on the angular resolution. Table \ref{EuclidGC} presents the detailed specifications for this mission. The number of detected galaxies by Euclid within a sky fraction $f_{\rm sky}=0.3636$ for a redshift bin of width $\Delta z$ around $\Bar{z}$ is given as,
\begin{equation}
 N(\Bar{z}) = 41253 ~\rm f_{\rm sky} ~\rm deg^{2} \int_{\Bar{z}-\rm \Delta z/2}^{\Bar{z}+\rm \Delta z/2} \dfrac{dN(z)/dz}{1~\rm deg^{2}} dz   
\end{equation}
Two nuisance parameters $\beta_{\rm 0}^{\rm Euclid}, \beta_{\rm 1}^{\rm Euclid}$ have been introduced in the galaxy bias factor detected by Euclid \citep[]{Audren:2012vy},
\begin{equation}
    b_{z} = \beta_{\rm 0}^{\rm Euclid}(1+z)^{0.5\rm \beta_{\rm 1}^{\rm Euclid}}
\end{equation}
As a prior we have chosen Gaussian priors with $\sigma=2.5\%$ for these $\beta$ parameters. 

\begin{table*}
\centering
\begin{tabular}{c|c|c|c|c|c}
\hline
    Parameter & $z_{\rm min}$ & $z_{\rm max}$ & $f_{\rm sky}$ & $\sigma_{z}$ & $\sigma_{\theta} ['']$ \\
\hline    
    Euclid & 0.7 & 2.0 & 0.3636 & 0.001(1+z) & 0 
\\    
\hline

\end{tabular}  
\caption{Euclid Galaxy Clustering specifications \citep[]{EuclidTheoryWorkingGroup:2012gxx}.} 
		\label{EuclidGC} 
 
\end{table*}

\subsubsection{SKA}
The most promising galaxy surveys for SKA are the SKA1-Mid Band1 (SKA1) and SKA1-Mid Band 2 (SKA2). Here we have considered the survey volume $S_{\rm area} = f_{\rm sky} \times 41253~ \rm deg^{2}$ according to the SKA baseline specifications \citep[]{Yahya:2014yva}. Since the Universe is expanding, the $\nu_{0} =$ 1420 MHz rest frame signal get redshifted with a observed frequency $\nu$ depending on at which redshift $z$ the transition has taken place, $ z = \dfrac{\nu_{0}}{\nu} - 1 $ and the error corresponding to redshift measurement is $\sigma_{z} = (1+z)^{2}\sigma_{\nu}$. The number of observed galaxies and the bias w.r.t the dark matter distribution have been taken from a simulation \citep[]{Yahya:2014yva} using the fitting formula,
\begin{equation}
    \dfrac{\rm dN(z)/\rm dz}{1~ \rm deg^{2}} = 10^{\rm c_{1}} z^{\rm c_{2}} \exp(-\rm c_{3}z)
\end{equation}
\begin{equation}
    b_{\rm HI}(z) = c_{4} ~\rm \exp(\rm c_{5}z)
\end{equation}
SKA1 has been divided into 64000 channels with bandwidth $\rm \delta \nu = 12.7 $~\rm kHz and bandwidth for SKA2 is $\rm \delta \nu = 12.8 $~\rm kHz. As in the case of Euclid, here also two nuisance parameters $\beta_{0}^{\rm SKA} ~\rm \& ~\rm \beta_{1}^{\rm SKA}$ have been introduced for the galaxy bias \citep[]{Sprenger:2018tdb},
\begin{equation}
    b_{z} = \beta_{0}^{\rm SKA}(1+z)^{\rm 0.5\beta_{1}^{\rm SKA}}
\end{equation}
with mean value 1 and corresponding 1$\sigma$ error is 0.025. In Table \ref{SKAGC}, we have outlined the instrumental specifications for SKA1 and SKA2.

\subsubsection{DESI}
The primary objective of Dark Energy Spectroscopic Instrument (DESI) \citep[]{2013arXiv1308.0847L}, \citep[]{2016arXiv161100036D} is to investigate the expansion history of the Universe. This is achieved by detecting the signature of Baryon Acoustic Oscillations present in cosmic structures and by measuring the growth rate of large-scale structures using redshift-space distortion measurements, which are influenced by peculiar velocities. However, in what follows, we will take SKA and Euclid as representative examples for constraining the parameters of the model and forecasting on the errors.  We keep the prospects of DESI reserved for Section \ref{cmb-mission}.

\begin{table*}
\centering
\begin{tabular}{c|c|c|c|c|c|c|c}
\hline
    Parameter & $\nu_{\rm min}$[\rm MHz] & $\nu_{\rm max}$[\rm MHz] & $z_{\rm min}$ & $z_{\rm max}$ & $S_{\rm area} [\rm deg^{2}]$ & $\delta_{\nu}$[\rm KHz] & $B$[\rm km] \\
\hline    
    SKA1  & 950 & 1760 & 0.00 & 0.5 & 5000   & 12.7 & 150 (5) \\
    SKA2  & 470 & 1290 & 0.10 & 2.0 & 30,000 & 12.8 & 3000 (5)
   \\ 
\hline
\end{tabular}  
\caption{SKA Galaxy Clustering specifications \citep[]{Yahya:2014yva}.}
		\label{SKAGC}
\end{table*}

\subsection{Upcoming Cosmic Shear Observations}

The matter distribution along the line of sight causes weak gravitational lensing of the observed galaxies, resulting in alignments of their images. A cosmic shear survey provides information about the matter distribution using auto and cross-correlations (at different redshift bins) of these alignments of the galaxies. The cosmic shear power spectrum of multipole $l$ at redshift bins $\{i,~j \}$ is given by \citep[]{Sprenger:2018tdb},
\begin{equation}
C_{\rm ij}^{l}= \dfrac{9}{16} \Omega_{m}^{2} H_{0}^{4} \int_{0}^{\infty} \dfrac{\rm dr}{r^{2}}  g_{i}(r)  g_{j}(r) P\left(k=\dfrac{l}{r},z(r)\right),
\end{equation}
where $P\left(k=\dfrac{l}{r},z(r)\right)$ is the three dimensional matter power spectrum. The function $g_{i}(r)$ denotes the convolution of the distribution of the observed galaxies with redshift error,
\begin{subequations}
\begin{align}
     g_{i}(r) &= 2r (1+z(r))\int_{r}^{\infty} dr' \dfrac{\eta_{i}(r'-r)}{r'}~, \label{eq:csa} \\
     \eta_{i}(r) &= H(r) ~\rm n_{i} ~\rm (z(r))~,\\
     n_{i}(z) &= \dfrac{D_{i}(z)}{\int_{0}^{\infty} D_{i}(z')dz'}~,\\
     D_{i}(z) &= \int_{z_{i}^{min}}^{z_{i}^{max}} \mathcal{P}(z,z')\dfrac{dn_{\rm gal}}{dz}(z')dz'~.
\end{align} 
\end{subequations}
As the intrinsic alignment of the galaxies are random, $C_{l}^{\rm ij}$ has contribution from an additional noise spectrum,
\begin{equation}
    N_{l}^{\rm ij} = \delta_{\rm ij} \sigma_{\rm shear}^{2}n_{i}^{-1},
\end{equation}
where the value of $\sigma_{\rm shear}=0.3$ is the root mean squared of the galaxy intrinsic ellipticity. $n_{i}$ is the number of galaxies in $i^{\rm th}$ redshift bin per steradian. The whole redshift range is divided into 10 redshift bins, thus, $n_{i} = \dfrac{n_{\rm gal}}{10} \times 3600 \left(\dfrac{180}{\pi}\right)^{2}$. 
   
\subsubsection{Euclid and SKA specifications}
We have summarized the specifications of the Euclid and SKA in Table \ref{CSspecs}. The unnormalized galaxy number density $dn_{\rm gal}/dz$ and its associated Gaussian error function $\mathcal{P}(z,z')$ are as follows,
\begin{equation}
    \dfrac{dn_{\rm gal}}{dz} = z^{\beta} \exp \left[-\left( \dfrac{z}{\alpha z_{m}}\right)^{\gamma}\right]
\end{equation}
\begin{equation}
    \mathcal{P}(z,z') = \dfrac{1}{\sqrt{2\pi}\sigma_{\rm photo-z}(1+z)} \exp\left[- \dfrac{(z-z')^{2}}{2\sigma_{\rm photo-z}^{2}(1+z)^2}\right]
\end{equation}
where $z$ and $z'$ are the true and measured redshifts.

\begin{table*}
\centering
\begin{tabular}{c|c|c|c|c|c|c|c|c|c|c|c}
\hline
    Experiments & $f_{\rm sky}$ & $n_{\rm gal} (\rm arcmin^{-2})$ & $z_{m}$ & $\alpha$ & $\beta$ & $\gamma$ & $f_{\rm spec-z}$ & $Z_{\rm spec-max}$ & $\sigma_{\rm photo-z}$ & $z_{\rm photo-max}$ & $\sigma_{\rm no-z}$\\
\hline    
    SKA1 & 0.1212   & 2.7 & 1.1 & $\sqrt{2}$ & 2 & 1.25 & 0.15 & 0.6 & 0.05 & 2.0 & 0.3                              \\
    SKA2 & 0.7272   & 10  & 1.3 & $\sqrt{2}$ & 2 & 1.25 & 0.5  & 2.0 & 0.03 & 2.0 & 0.3                      \\
    Euclid & 0.3636 & 30  & 0.9 & $\sqrt{2}$ & 2 & 1.5  & 0.0  & 0.0 & 0.05 & 4.0 & 0.3
\\
\hline

\end{tabular}  
\caption{Cosmic Shear specifications for SKA \citep[]{Harrison:2016stv} and Euclid \citep[]{Audren:2012vy}.}
		\label{CSspecs}
 
\end{table*}

\section{Fisher Forecast, MCMC Analysis and Errors}
Having set the stage, let us now  briefly discuss the Fisher matrix forecast method, Markov Chain Monte Carlo (MCMC) analysis and also the theoretical and realistic errors in the upcoming experiments taken into account in our analysis. Fisher matrix method is an efficient tool to estimate accuracy at which we can constrain a model parameter for an upcoming experiment  \citep[]{Verde:2009tu}, \citep[]{2009arXiv0906.4123C}. The Fisher matrix $F_{ ij}$ is defined as the second derivative of the log likelihood $\ln \mathcal{L}$ or the chi-squared function $\chi^{2}$ with respect to the parameters of interest, evaluated at their best-fit values or fiducial values, 
\begin{equation}\label{Fij}
    F_{ ij}=-\left\langle \dfrac{\partial^{2}\ln\mathcal{L}}{\partial q_{i} \partial q_{j}} \right\rangle = - \dfrac{\partial^{2}\ln\mathcal{L}}{\partial q_{i} \partial q_{j}}~\Bigg|_{q_{0}}.
    \end{equation}

It basically approximates the logarithm of the likelihood function as a multivariate Gaussian function of the parameters $q$ at their fiducial values $q_{0}$. The inverse of the Fisher matrix is the covariance matrix,
\begin{equation}
 {\rm Cov}(q_{i},q_{j}) \geq [F^{-1}]_{ij}   . 
\end{equation}
   
The diagonal elements of the Fisher matrix give the square of the 1$\sigma$ bounds on the corresponding cosmological parameters, 

\begin{equation}
 \sigma(\alpha_{i})=\sqrt{[F^{-1}]_{ii} }   
\end{equation}

As mentioned before, to solve the Boltzmann equations in the DM-neutrino interaction scenario, we have used the modified version \citep[]{Mosbech:2020ahp} of the publicly available code \texttt{CLASS} 
\citep[]{Blas_2011}. We have estimated the Fisher matrix using the publicly available code \texttt{MontePython} v3.4 \citep[]{Audren:2012wb}, \citep[]{Brinckmann:2018cvx} using the likelihood of the corresponding future experiments.

Further,  with the modified version of \texttt{CLASS} \citep[]{Blas_2011}, we have performed MCMC technique to analyze the errors and correlations of the model parameters using the mock data generated from \texttt{MontePython} v3.4 code \citep[]{Audren:2012wb}, \citep[]{Brinckmann:2018cvx}  for upcoming 21-cm and LSS missions. In contrast to the customary Fisher approach where the analysis is solely dependent on the fiducial values of the parameters and it assumes log-likelihood function as Gaussian function, the Bayesian MCMC method provides the capability to investigate both Gaussian and non-Gaussian posteriors and offers immunity against potential numerical stability challenges associated with selecting the step size for numerical derivatives. 

We have carried out the Fisher forecast method and Bayesian MCMC techniques to investigate the possible $1 \sigma$ errors on the (6+2)- parameter scenario in the present context.  The relevant parameters are the baseline 6 parameters  $w_{b}, ~w_{\rm nudm}, ~\ln[10^{10}A_{s}], ~n_{s}, ~H_{0}, ~\tau_{\rm reio}$ along with DM-neutrino interaction strength `$u$' and sum of neutrino masses `$M_{\rm tot}$' using forthcoming experiments SKA and Euclid.

Further, in our work we have considered a large scale cut-off above which the small-angle approximation is not valid. For this we have taken $k_{\rm min}=0.02$ \rm Mpc$^{-1}$ which cuts off all the scales above the assumed redshift bin width of the corresponding experiment. On small scales we have adopted two approaches: 
\begin{itemize} 
\item One is the conservative approach \citep[]{Sprenger:2018tdb} where a redshift dependent non-linear cut-off $k_{\rm NL}(z) = k_{\rm NL}(0)(1+z)^{2/(2+n_{s})}$ is being introduced, here $n_s$ is the scalar spectral index. In our analysis, we have taken into account $k_{\rm NL}(0)=0.2 \rm h ~Mpc^{-1}$ for Euclid, SKA Galaxy Clustering and SKA Intensity Mapping experiments. For Euclid and SKA Cosmic Shear experiments we have implied multiple limits starting from $l_{\rm min}=5$ to redshift dependent non-linear cut-off \citep[]{Smith:2002dz} $l_{\rm max}^{i} = k_{\rm NL}(z)\times \bar{r}_{\rm peak}^{i} $, where $k_{\rm NL}(0)=0.5 \rm h ~Mpc^{-1}$ is being considered for conservative approach. Here $\bar{r}_{\rm peak}^{i}$ is defined as $\bar{r}_{\rm peak}^{i}=(\sum_{j>i} \dfrac{\int_{0}^{\infty} dr.r/r^{2} g_{i}(r) g_{j}(r)}{dr / r^{2} g_{i}(r) g_{j}(r)})/(N-1)$, where $N$ is the number of redshift bins and $g_{i}(r)$ denotes the convolution of the distribution of the observed galaxies and it is defined in equation \eqref{eq:csa}. For MCMC analysis, we have considered conservative approach only.

\item The second one is rather a realistic approach \citep[]{Sprenger:2018tdb} where we have chosen the theoretical error $k_{\rm max}=10 \rm h ~Mpc^{-1}$ for Euclid and SKA Galaxy Clustering and SKA Intensity Mapping experiments. For the Euclid and SKA Cosmic Shear experiments, we have incorporated a realistic error approach by considering  $k_{\rm NL}(0) = 2.0 \rm h ~Mpc^{-1}$. For Fisher forecast, we have taken this realistic approach parallel to the conservative one.
\end{itemize}

\begin{table*}
\centering
\begin{tabular}{||c c c c c c c c c c c ||} 
 \hline
 Expts. & CS & GC & $\sigma(w_{b})$ & $\sigma(w_{\rm nudm})$ & $\sigma(\ln[10^{10}A_{s}])$ & $\sigma(n_{s})$ & $\frac{\sigma(H_{0})}{\left[\frac{\rm km}{s \rm Mpc}\right]}$ & $\sigma(\tau_{\rm reio})$ &$\sigma(u)$ & $\frac{\sigma(M_{\rm tot})}{\rm eV}$  \\ [0.5ex] 
 \hline\hline
 Planck18+SKA1 & c & - & 0.00013 & 0.00078 & 0.01196 & 0.00306 & 1.147 & 0.00604 & $1.293 \times 10^{-6}$ & 0.1172  \\ 

 - & r & - & 0.00015 & 0.00114 & 0.008245 & 0.00339 & 1.881 & 0.00471 & $8.114 \times 10^{-7}$ & 0.1563 \\

 - & - & c & 0.00013 & 0.00091 & 0.01102 & 0.00316 & 0.630 & 0.00607 & $1.348\times 10^{-6}$ & 0.0833 \\

 - & c & c & 0.00012 & 0.00061 & 0.01083 & 0.00284 & 0.610 & 0.00581 & $1.044\times 10^{-6}$ & 0.0715 \\

 - & r & c & 0.00013 & 0.00049 & 0.00808 & 0.00277 & 0.536 & 0.00419 & $7.815 \times 10^{-7}$ & 0.0452 \\

- & - & r & 0.00012 & 0.00045 & 0.00688 & 0.00271 & 0.226 & 0.00371 & $2.594 \times 10^{-7}$ & 0.0234\\

 - & c & r & 0.00012 & 0.00041 & 0.00787 & 0.002595 & 0.081 & 0.00423 & $2.482 \times 10^{-7}$ & 0.0227 \\

 - & r & r & 0.00012 & 0.00039 & 0.00785 & 0.00258 & 0.225 & 0.00423 & $2.400 \times 10^{-7}$ & 0.0212  \\
 \hline\hline
 Planck18+SKA2 & c & - & 0.00012 & 0.000393 & 0.01022 & 0.00257 & 0.471 & 0.00546 & $2.033 \times 10^{-7}$ & 0.0487\\

 - & r & - & 0.00012 & 0.00035 & 0.00694  & 0.00204 & 0.314 & 0.00369 & $9.083 \times 10^{-8}$ & 0.0251 \\

- & - & c & 0.00011 & 0.00028 & 0.00593 & 0.00166 & 0.085 & 0.00319 & $4.775 \times 10^{-8}$ & 0.0132 \\

- & c & c & 0.00011 & 0.00057 & 0.013699 & 0.00226 & 0.118 & 0.00771 & $ 4.805 \times 10^{-8}$ & 0.0246\\

- & r & c & 0.00011 & 0.00028 & 0.00477 & 0.00139 & 0.089 & 0.00282 & $ 3.582 \times 10^{-8}$ & 0.0086\\

- & - & r & 0.00011 & 0.00030 & 0.00507 & 0.00075 & 0.090 & 0.00297 & $ 1.636 \times 10^{-6}$ & 0.0096 \\

- & c & r & 0.00013 & 0.00025 & 0.00443 & 0.00065 & 0.091 & 0.00234 & $1.551 \times 10^{-8}$ & 0.0095 \\

- & r & r & 0.00020 & 0.00073 & 0.01008 & 0.00091 & 0.183 & 0.00638 & $1.557 \times 10^{-8}$ & 0.0111 \\

\hline\hline

Planck18+Euclid & c & - & 0.00013 & 0.00043 & 0.00856 & 0.00276 & 0.764 & 0.00449 & $8.553 \times 10^{-7}$ & 0.0716\\

- & r & - & 0.00012 & 0.00038 & 0.01019 & 0.00265 & 0.365 & 0.00548 & $ 2.529 \times 10^{-7}$ & 0.0377 \\

- & - & c & 0.00012 & 0.00037 & 0.006496 & 0.00241 & 0.197 & 0.00334 & $8.313 \times 10^{-8}$ & 0.0211 \\

- & c & c & 0.00012 & 0.00029 &  0.00623 & 0.002398 & 0.186 & 0.00343 & $ 8.088 \times 10^{-8}$ & 0.0181\\

- & r & c & 0.00011 & 0.00029 & 0.00668 & 0.00214 & 0.174 & 0.00365 & $ 7.234 \times 10^{-8}$ & 0.0181\\

- & - & r & 0.00011 & 0.00039 & 0.00758 & 0.00139 & 0.166 & 0.00406 & $2.434 \times 10^{-8}$ & 0.0179\\

- & c & r & 0.00012 & 0.00025 & 0.00652 & 0.00112 & 0.125 & 0.00344 & $2.340 \times 10^{-8}$ & 0.0163\\

- & r & r & 0.00011 & 0.00027 & 0.00636 & 0.00108 & 0.126 & 0.00346 & $ 2.317\times 10^{-8}$ & 0.0151\\
\hline\hline
- +SKA1 IM2 (c)& c & c & 0.00011 & 0.00028 & 0.00682 & 0.00224 & 0.106 & 0.00375 & $7.905\times 10^{-8}$ & 0.0159 \\

- +SKA1 IM2 (r)& r & r & 0.00010 & 0.00011 & 0.00550 & 0.00105 & 0.083 & 0.00310 & $2.322 \times 10^{-8}$ & 0.0112 \\
 \hline\hline

 Planck18+SKA1 & IM1 & c & 0.00012 & 0.00029 & 0.00724 & 0.00236 & 0.104 & 0.00399 & $ 3.865 \times 10^{-8}$ & 0.0155 \\
Planck18+SKA1 & IM1 & r & 0.00012 & 0.00030 & 0.00725 & 0.00236 & 0.101 & 0.00401 & $ 3.800 \times 10^{-8}$ & 0.0152 \\

 Planck18+SKA1 & IM2 & c & 0.00012 & 0.00042 & 0.01036 & 0.00266 & 0.204 & 0.00551 & $7.811 \times 10^{-7}$ & 0.0353 \\

Planck18+SKA1 & IM2 & r & 0.00012 & 0.00032 & 0.00825 & 0.00242 & 0.114 & 0.00438 & $ 3.494 \times 10^{-7}$ & 0.0182 \\

Planck18+SKA1 & IM1,IM2 & c & 0.00012 & 0.00029 & 0.00719 & 0.00228 & 0.096 & 0.00395 & $3.628 \times 10^{-8}$ & 0.0154 \\
 Planck18+SKA1 & IM1,IM2 & r & 0.00011 & 0.00031 & 0.00785 & 0.00211 & 0.081 & 0.00427 & $ 3.475\times 10^{-8}$ & 0.0148 \\
 \hline\hline

\end{tabular}
\caption{The expected 1$\sigma$ uncertainties on the cosmological parameters for various future experiments SKA1 Galaxy Clustering (GC) and Cosmic Shear (CS), SKA2 GC $\&$ CS, Euclid GC $\&$ CS, Euclid + SKA1 Intensity Mapping (IM) Band 2 GC $\&$ CS and SKA1 IM1 $\&$ IM2 GC and CS. Here `c' and `r' represents the conservative and realistic theoretical error for corresponding future experiments.}
\label{error1}
\end{table*}

\section{Results and Analysis}
We now present a comparative status of the results obtained from different types of analysis (Fisher forecast, MCMC) for different future missions  and their possible combinations for interacting DM-massive neutrino scenario. In Table \ref{error1} we have presented the 1$\sigma$ error on the model parameters for different Galaxy clustering, Cosmic Shear and Intensity Mapping experiments. While estimating errors using Fisher matrix forecast method, we have taken into account Planck18 as the default CMB experiment so as to remain consistent with latest observational constraints from CMB \citep[]{Mosbech:2020ahp}. However, for our forecast analysis, we have used the mock catalogue of Planck18 which has been generated from \texttt{MontePython} v3.4  code \citep[]{Audren:2012wb}, \citep[]{Brinckmann:2018cvx} instead of Planck18 real data. This is done with the sole intention of reducing the computational time, without losing any major information from real data, thereby respecting the latest constraints of Planck18. Further, for all the forecasts and MCMC analysis we have chosen the fiducial values of our model parameter as $w_\text{b}=0.02237,~w_\text{\rm nudm}=0.12010,~\ln[10^{10}A_\text{s}]=3.0447,~n_\text{s}=0.9659,~H_\text{0}=67.8,~\tau_\text{\rm reio}=0.0543,~ u=0, ~M_\text{\rm tot}=0.06$ eV where we have assumed a minimum value for total neutrino mass and interaction parameter, other parameter values are based on Planck18 data \citep[]{refId0}. 

Below we will analyze the corresponding errors on the model parameters `$u$' that represents the DM-massive neutrino interaction and the total neutrino mass parameter `$M_{\rm tot}$'. However, for MCMC, we will find out constraints on the usual 6 parameters on top of these 2 parameters, and investigate their correlations, if any.

\begin{figure}
	\includegraphics[width=\columnwidth]{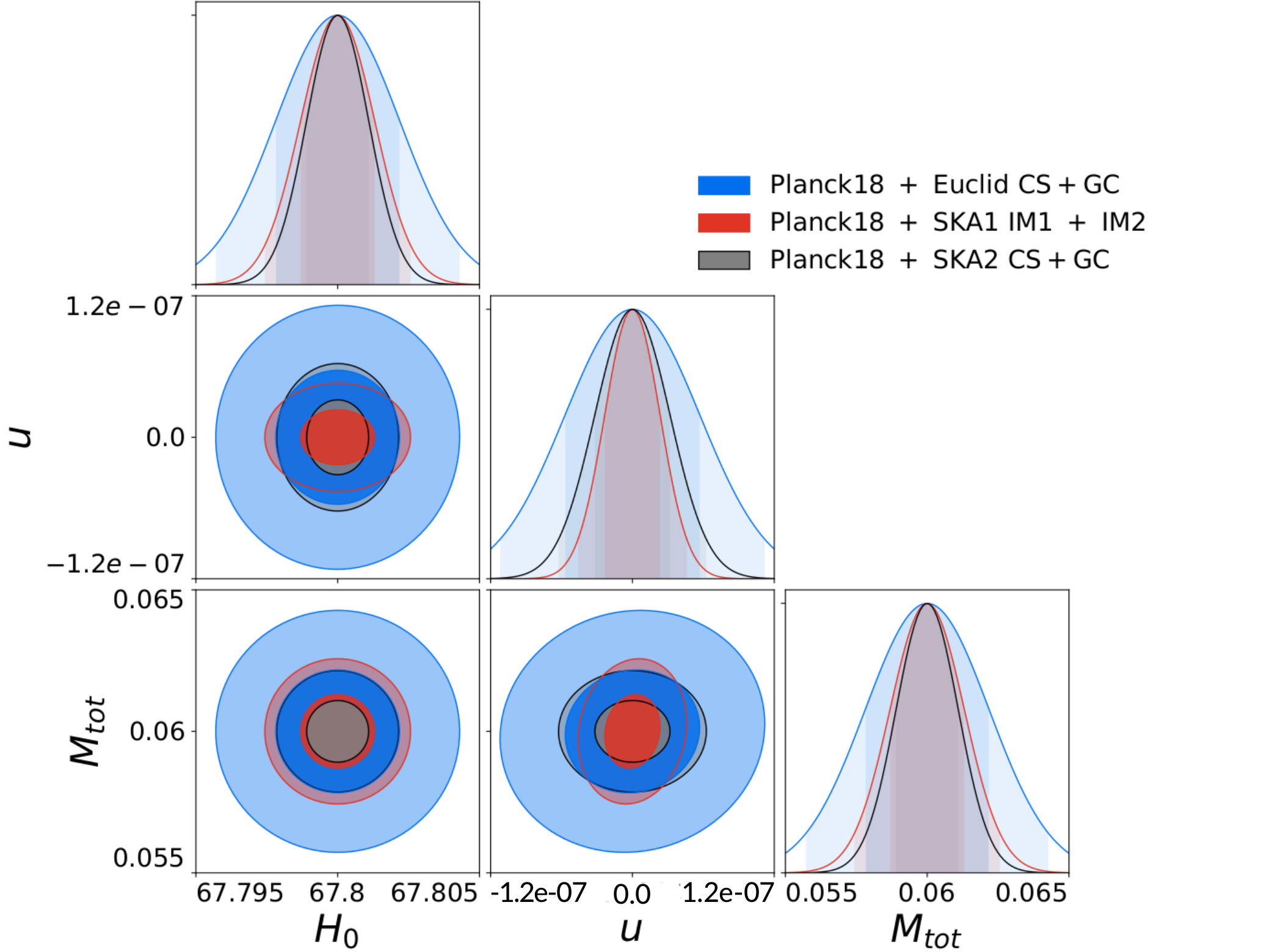}
	\caption{From Fisher matrix analysis the above figure shows marginalized 1$\sigma$ and 2$\sigma$ contours and one-dimensional posteriors for the cosmological parameters using future experiments Planck18 + Euclid (CS + GC), Planck18 + SKA1 Intensity Mapping Band (1 + 2), Planck18 + SKA2 (CS + GC). For the above figure the analysis is being done only for the Conservative approach of the theoretical error. }
\label{LSS_CONS}
\end{figure}

\begin{figure}
	\includegraphics[width=\columnwidth]{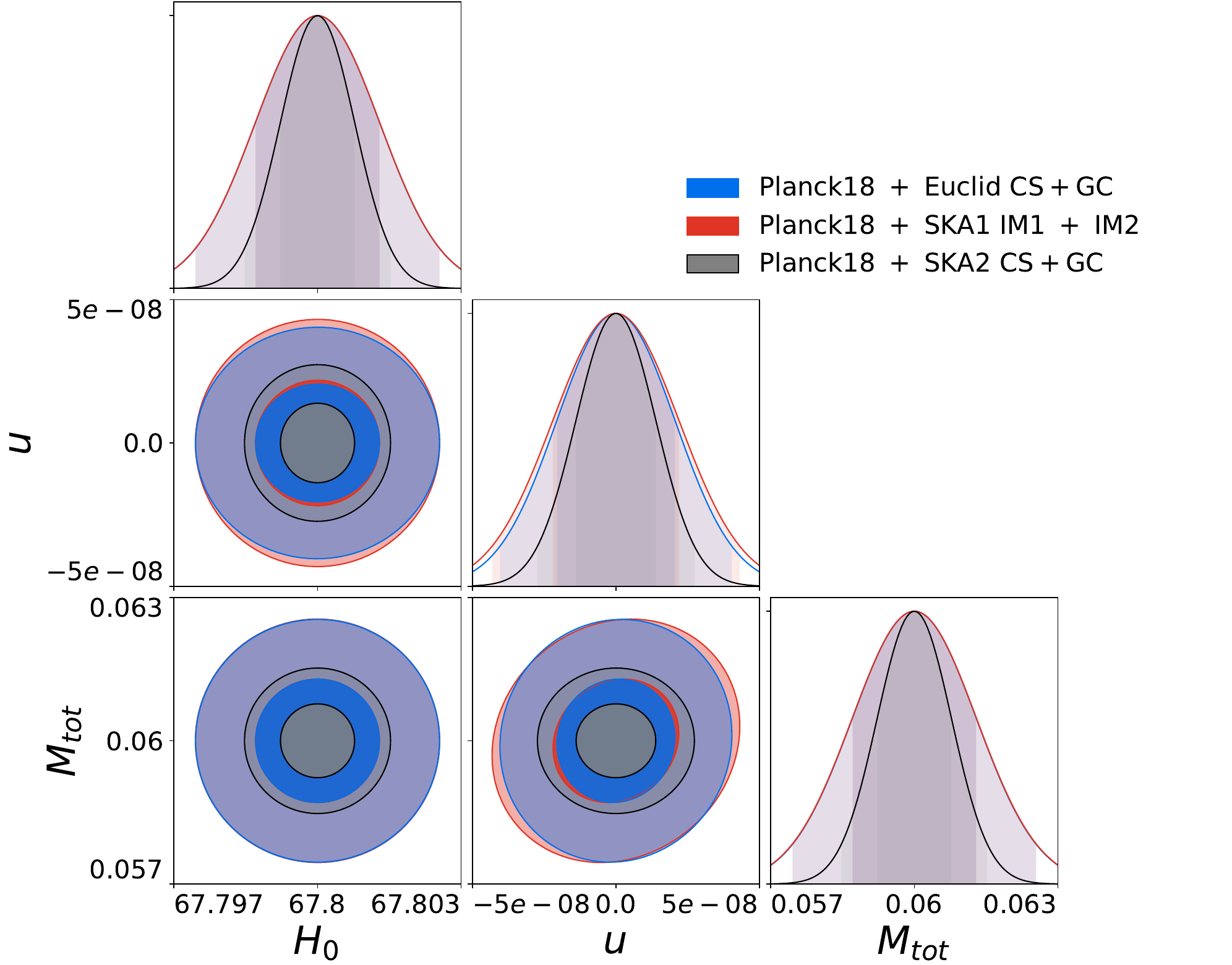}
	\caption{ From Fisher matrix analysis the above figure shows marginalized 1$\sigma$ and 2$\sigma$ contours and one-dimensional posteriors for the cosmological parameters using future experiments Planck18 + Euclid (CS + GC), Planck18 + SKA1 Intensity Mapping Band (1 + 2), Planck18 + SKA2 (CS + GC). For the above figure the analysis is being done only for the Realistic approach of the theoretical error.}
\label{LSS_Real}
\end{figure}

\begin{figure}
	\includegraphics[width=\columnwidth]{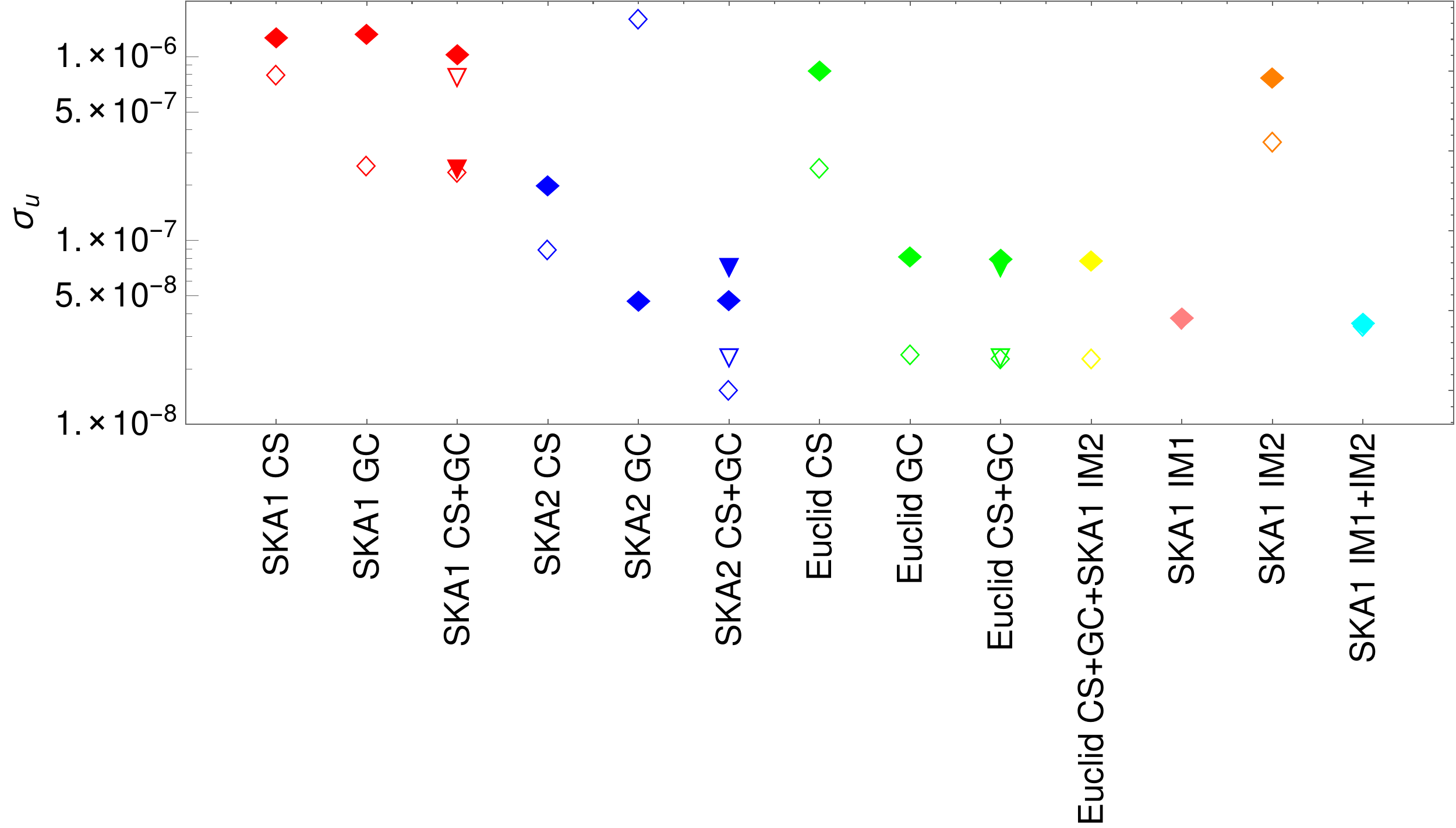}
	\caption{The above figure depicted 1$\sigma$ uncertainties on the DM-massive neutrino interaction parameter `$u$' for various cosmological experiments including Planck18 to the baseline cosmological parameter. Here different symbols $\blacklozenge, \diamondsuit$  represents Conservative error, Realistic error of the single future experiment, $\blacktriangledown$ indicates the Conservative error of the first experiment and Realistic error of the second experiment and $\triangledown$ specifies the Realistic error of the first experiment and Conservative error of the second experiment.}
\label{sigma_u_LSS}
\end{figure}

\begin{figure*}
\includegraphics[scale=0.237]{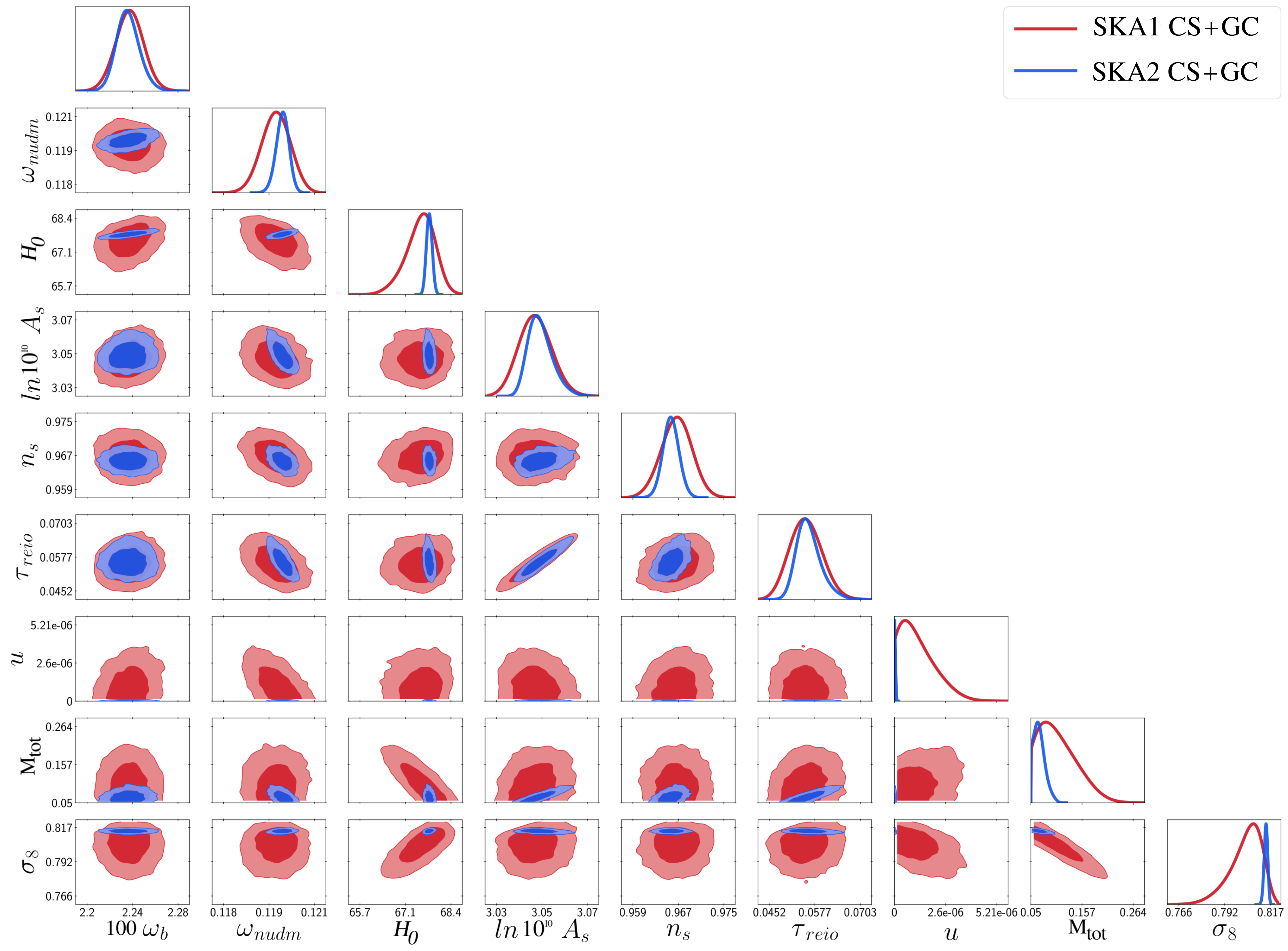}
\caption{ From MCMC techniques the above figure depicts the marginalized 1$\sigma$ and 2$\sigma$ contours and one-dimensional posteriors for the cosmological parameters using future experiments Planck18 + SKA1 (CS + GC), Planck18 + SKA2 (CS + GC). For the above figure the analysis is being done only for the Conservative approach of the theoretical error.}
\label{SKA1_GC+CS-vs-SKA2_GC+CS_triangle}
\end{figure*}

\begin{figure*}
\includegraphics[scale=0.237]{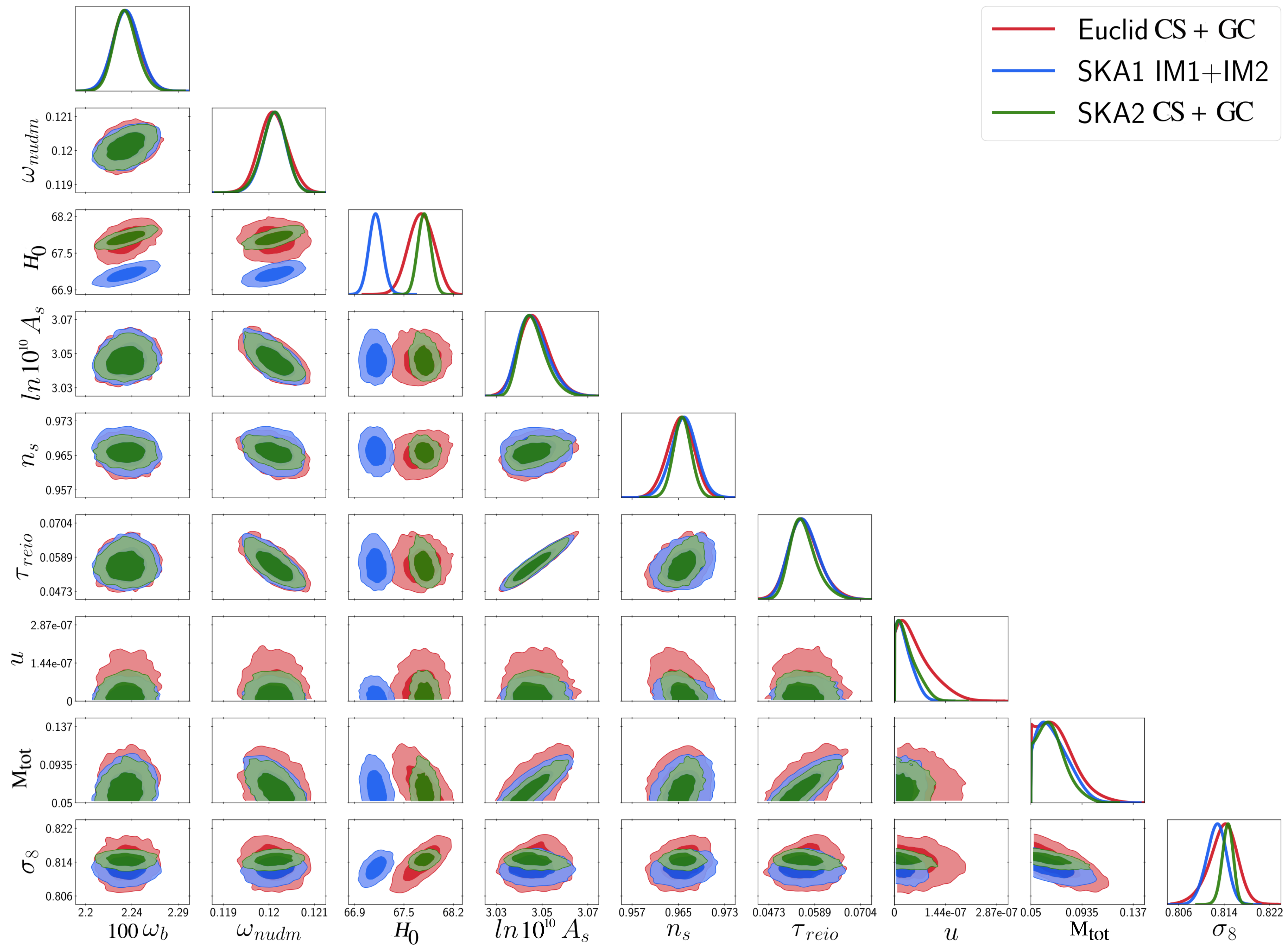}
\caption{ From MCMC techniques the above figure depicts the marginalized 1$\sigma$ and 2$\sigma$ contours and one-dimensional posteriors for the cosmological parameters using future experiments Planck18 + Euclid (CS + GC), Planck18 + SKA1 Intensity Mapping Band (1 + 2), Planck18 + SKA2 (CS + GC). For the above figure the analysis is being done only for the Conservative approach of the theoretical error.}
\label{Euclid_GC+CS-vs-SKA1_IM1+IM2-vs-SKA2_GC+CS_triangle}
\end{figure*}

\subsection{Constraints on the interaction strength `$u$'}

\begin{itemize}
    \item The error ellipses from the Fisher matrix analysis have been plotted in Figure \ref{LSS_CONS} $\&$ \ref{LSS_Real} for the conservative and realistic approaches to the theoretical error respectively.   From Table \ref{error1} it is evident that the upper bound on the interaction parameter `$u$' is $u \leq 1.551 \times 10^{−8}$ 
    from SKA2 CS + GC future experiments, where conservative error in CS and realistic error in GC experiment have been considered. While for other combinations of SKA2 CS $\&$ GC experiments, we obtain much weaker constraints on the model parameters. The SKA1 CS + GC experiments demonstrate significantly lower constraints on the model parameters, approximately one magnitude weaker than the SKA2 experiments. Conversely, the SKA1 IM experiments provide constraints on the model parameters similar to those obtained with SKA2. The sensitivity of the model parameters solely depends on the instrumental specifications of the corresponding experiments. The SKA2 and SKA1 IM experiments cover a much broader redshift range and larger sky fraction than the SKA1 mission. As a result, the SKA1 mission is unable to effectively constrain the model parameters, in contrast to the SKA2 and SKA1 IM experiments.
   It should also be noted that in Figure \ref{LSS_CONS} $\&$ \ref{LSS_Real}, although the negative value of `$u$' are shown, this is basically an artifact of considering zero mean for the parameter. In reality, a negative interaction strength holds no physical significance as such, as has been rightly found out from MCMC analysis  by non-negative bounds for `$u$' in Figure \ref{SKA1_GC+CS-vs-SKA2_GC+CS_triangle} $\&$ \ref{Euclid_GC+CS-vs-SKA1_IM1+IM2-vs-SKA2_GC+CS_triangle}.

    \item The other LSS experiment Euclid CS + GC which has been launched recently, also puts stronger upper bounds on the interaction parameter `$u$', which is $u \leq 10^{-8}$ that conforms with the constraints from SKA2. Compared to the SKA1 mission, Euclid's galaxy cluster and cosmic shear experiments have undergone significant improvements in their instrumental specifications, these enhancements have resulted in increased sensitivity to the DM-massive neutrino interaction parameter. In other words, Euclid's experiments are better equipped to detect and constrain the interaction between DM and neutrinos. This improvement in sensitivity allows Euclid to provide more robust and stringent bounds on the interaction parameter compared to the SKA1 mission.

     \item In Figure \ref{sigma_u_LSS} we have  compared the 1$\sigma$ errors on `$u$' as obtained from different experiments and their combinations. For this we have made use of both conservative and realistic approaches. From this figure it is also evident that SKA2 CS~+~GC has the maximum sensitivity on `$u$'. Hence it would be able to constrain the interaction parameter better than other experiments under consideration.

   \item Figure \ref{SKA1_GC+CS-vs-SKA2_GC+CS_triangle}  depicts the comparative analysis of the results for MCMC for the SKA1 CS~+~GC and SKA2 CS~+~GC experiments. It represents the marginalized 1$\sigma$ and 2$\sigma$ contours and one-dimensional posteriors for all the cosmological parameters for the model under consideration. An immediate conclusion that transpires from the confidence contours is  that SKA2 would be able to constrain all the $(6+2)$ cosmological parameters much tighter than the corresponding constraints from SKA1. This is mostly because 
   of the fact that the sensitivity, redshift range, bin-width and sky fraction of SKA2 are significantly improved compared to the SKA1 experiment. Further, although SKA2 reduces the 1$\sigma$ and $2\sigma$ contours for $H_{0}$ and $\sigma_{8}$ parameters compared to SKA1 and also from the current constraints from Planck18, the mean values of both  $H_{0}$ and $\sigma_{8}$ from SKA1 and SKA2 both are in agreement with the Planck18 results. As a result, the scenario may not be much useful if one wants to address tensions for either of the two parameters, at least for post-reionization epoch. However, in this method, inherently we expect a slight bias towards the Planck values as the mock data is generated from Planck best-fit values. We need to wait for the real data in order to make any conclusive comment.

   \item In order to investigate the potential of the Intensity Mapping experiments in constraining the parameters, in Figure \ref{Euclid_GC+CS-vs-SKA1_IM1+IM2-vs-SKA2_GC+CS_triangle} we have compared the MCMC results from future experiments Euclid CS~+~GC, SKA1 IM 1~$\&$~2 and SKA2 CS~+~GC. The 1$\sigma$ and 2$\sigma$ confidence contours reveal that the SKA1 IM1 experiment and combination of Euclid and SKA1 IM2 and SKA1 IM1 with SKA1 IM2 also give similar constraints on `$u$' as in the case of Euclid. However, SKA1 IM2 alone gives much weaker constraint on `$u$'. Among these three future experiments, SKA1 IM and SKA2 would constrain the cosmological parameters more stringently compared to Euclid. In these plots the posteriors of $H_{0}$ for SKA1 IM experiment is different from posteriors of SKA2 $\&$ Euclid, but the range of $H_{0}$ for SKA1 IM experiment is within the error bar of Planck18.

\end{itemize}

\begin{figure}
	\includegraphics[width=\columnwidth]{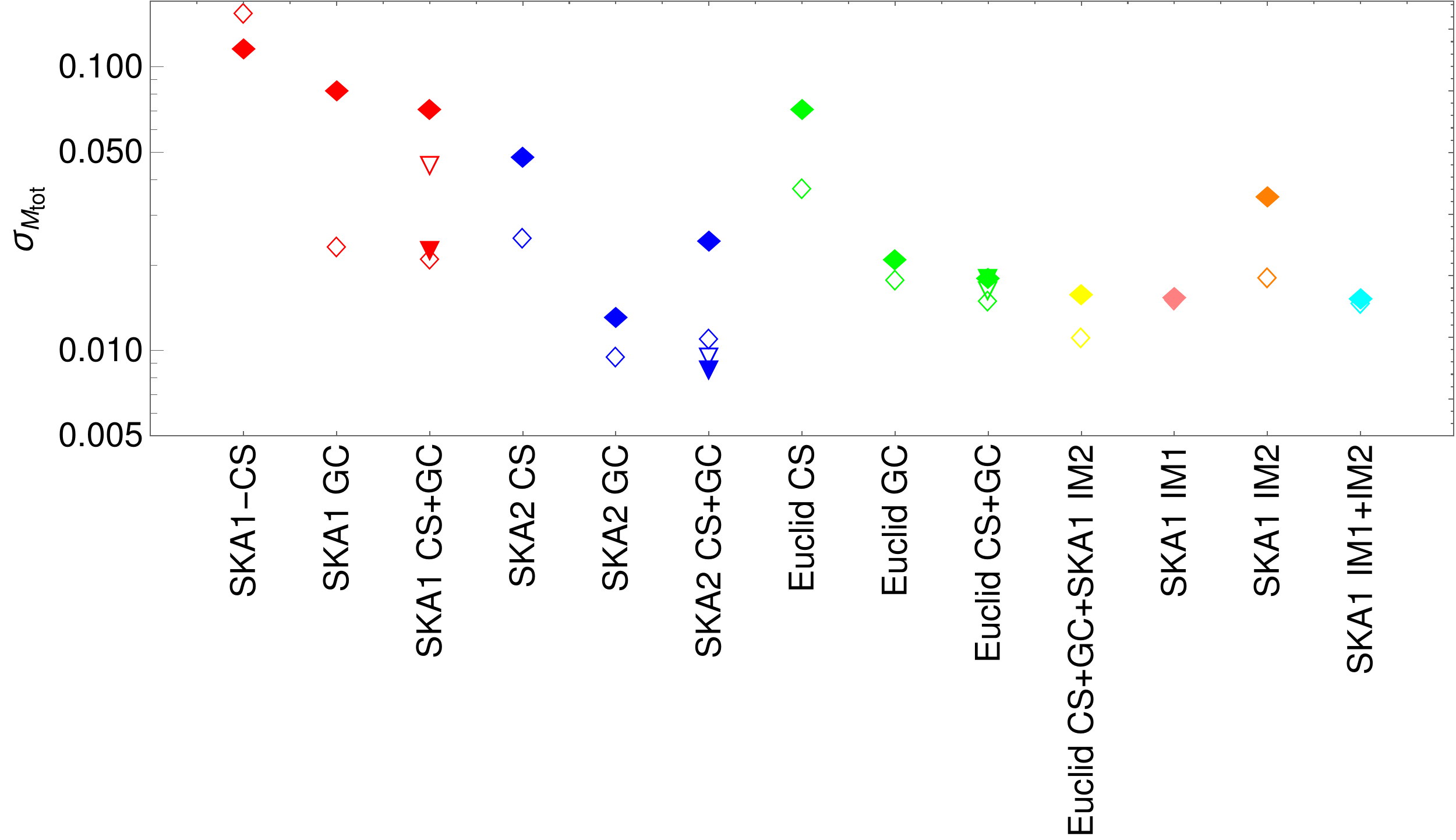}
	\caption{The above figure depicted 1$\sigma$ uncertainties on the mass of neutrinos for various cosmological experiments including Planck18 to the baseline cosmological parameter. Here different symbols $\blacklozenge, \diamondsuit$  represents Conservative error, Realistic error of the single future experiment, $\blacktriangledown$ indicates the Conservative error of the first experiment and Realistic error of the second experiment and $\triangledown$ specifies the Realistic error of the first experiment and Conservative error of the second experiment. }
\label{sigma_mtot_new_LSS}
\end{figure}

\subsection{Constraints on the total neutrino mass `$M_{\rm tot}$'}
\begin{itemize}
    \item Experimental evidence from neutrino oscillation experiments \citep[]{Cleveland_1998}, \citep[]{PhysRevLett.81.1562}, \citep[]{PhysRevLett.89.011301} confirms that at least two neutrino species possess mass, which was suggested in \citep[]{Pontecorvo:1967fh}, \citep[]{Gribov:1968kq} previously. However, the precise magnitude of their masses and their mass hierarchy remains undetermined, with no conclusive findings in the realms of cosmology \citep[]{Gariazzo:2018pei} or $\beta$-decay experiments \citep[]{KATRIN:2021uub}.
 One of the most remarkable accomplishments anticipated from the Euclid and SKA missions is the precise determination of the sum of neutrino masses, which may as well help us determine the mass hierarchy of neutrinos. In our analysis, for simplicity, we have considered a degenerate mass `$M_{\rm tot}$' for neutrinos. The strongest $1\sigma$ bound on `$M_{\rm tot}$' is $8.63 \times 10^{-3}$ eV  from SKA2 CS + GC future experiments, where conservative error in GC and realistic error in CS experiment have been taken into account, whereas other combination in SKA2 CS $\&$ GC experiments we get much weaker constraints on this model parameter. 
    \item The other LSS future experiments SKA1, Euclid, SKA1 IM constrains the `$M_{\rm tot}$' parameter weakly compared to the SKA2 experiment. In Figure \ref{sigma_mtot_new_LSS} we have indicated the 1$\sigma$ error on the `$M_{\rm tot}$' parameter. The figure provides compelling evidence that the SKA2 (CS + GC) achieves the highest level of accuracy in estimating the `$M_{\rm tot}$' parameter, boasting the smallest error among all the considered measurements.
\end{itemize}

\subsection{Dependence of forecasted error on fiducial values}
\begin{figure}
	\includegraphics[width=\columnwidth]{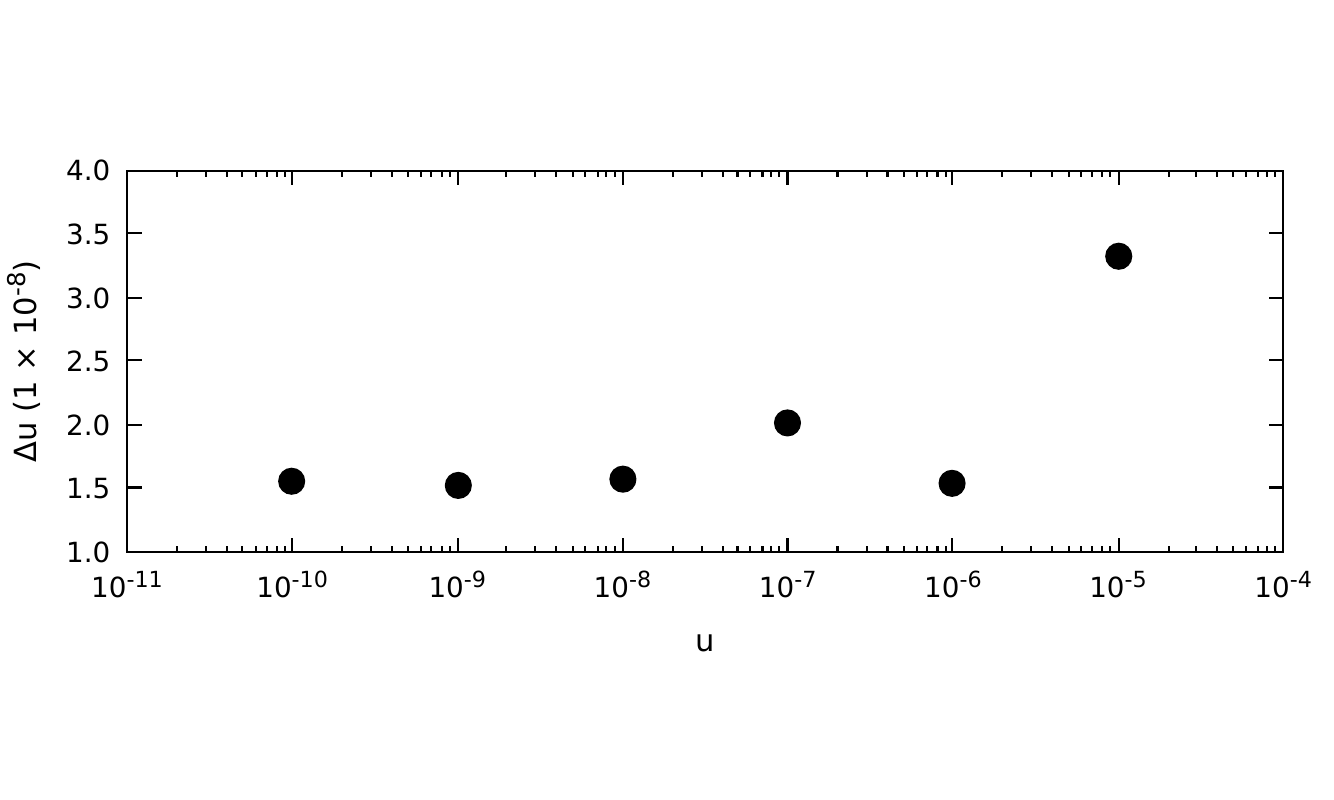}
	\caption{Dependence of Fisher forecasted error on the mean value of u for SKA2 (CS + GC) experiments.}
\label{forecast_error_u}
\end{figure}

In order to overcome any fiducial-dependent bias on the above-mentioned results, we have further examined the dependence, if any, of the choice of fiducial values of the parameters `$u$' and `$M_{\rm tot}$' on the predicted error $\sigma(u)$ and $\sigma(M_{\rm tot})$ for future  experiments. For this, we have plotted the estimated error $\Delta u$ on the mean value of `$u$' in Figure \ref{forecast_error_u} for SKA2 (CS + GC) experiments. We have found that, even though there is slight variation of the estimated error as we take different fiducial values, the order of magnitude of the errors on $\sigma(u)$ and $\sigma(M_{\rm tot})$ are same for the range of mean values considered in our work.  In addition, the `$u$' and `$M_{\rm tot}$' error forecasts demonstrate a limited reliance on the selection of mean values within our target range, with deviations mostly confined to the sub-percent range. This slight change in the predicted error can be attributed to systematics, statistical uncertainties as well as intrinsic randomness in the Fisher forecast and MCMC processes. This exercise leads us to believe that the estimated errors do not have any significant dependence on the choice of the fiducials. Our results are thus quite generic for the experiments under consideration.

\begin{table*}
\centering
\begin{tabular}{||c c c c c c c c c ||} 
 \hline
 Expts. & $\sigma(w_{b})$ & $\sigma(w_{\rm nudm})$ & $\sigma(\ln[10^{10}A_{s}])$ & $\sigma(n_{s})$ & $\frac{\sigma(H_{0})}{\left[\frac{\rm km}{s \rm Mpc}\right]}$ & $\sigma(\tau_{\rm reio})$ &$\sigma(u)$ & $\frac{\sigma(M_{\rm tot})}{\rm eV}$  \\ [0.5ex] 
 \hline\hline

 LiteBIRD & 0.00018 & 0.00116 & 0.00500 & 0.00446  & 1.848 & 0.00208 & $3.048 \times 10^{-6}$ & 0.1808  \\ 
 COrE-M5  & 0.00014 & 0.00210 & 0.00315 & 0.00182 & 1.891 & 0.00315 & $1.279 \times 10^{-7}$ & 0.1134 \\
 CMB-S4   & 0.00003 & 0.00038 & 0.00487 & 0.00170  & 0.261 & 0.00279 & $4.613 \times 10^{-8}$ & 0.0203 \\
 PICO     & 0.00007  & 0.00105 & 0.02075 & 0.00167 & 1.287 & 0.00894 & $1.599 \times 10^{-7}$ & 0.0961 \\
 LiteBIRD + CMB-S4 & 0.00006 & 0.00130 & 0.00785 & 0.00157 & 1.391 & 0.00302 & $7.619 \times 10^{-8}$ & 0.0983 \\
 COrE-M5 + CMB-S4 & 0.00006 & 0.00124 & 0.00730 & 0.00156 & 1.313 & 0.00280 & $7.388 \times 10^{-8}$ & 0.0924 \\
 CMB-S4 + DESI & 0.00003 & 0.00034 & 0.00702 & 0.00157 & 0.321 & 0.00379 & $4.804 \times 10^{-8}$ & 0.0331 \\
 CMB-S4 + Euclid + SKA1 IM2 & 0.00003 & 0.00009 & 0.00334 & 0.00071 & 0.039 & 0.00181 & $1.465 \times 10^{-8}$ & 0.0060 \\ 
 CMB-S4 + Euclid CS + GC & 0.00002 & 0.00021 & 0.00508 & 0.00086 & 0.098 & 0.00286 & $1.583 \times 10^{-8}$ & 0.0109 \\
 CMB-S4 + SKA1 CS + GC & 0.00003 & 0.00037 & 0.00541 & 0.00183 & 0.408 & 0.00332 & $4.208 \times 10^{-8}$ & 0.0334 \\
 CMB-S4 + SKA2 CS + GC & 0.00004 & 0.00070 & 0.01789 & 0.00182 & 0.121 & 0.01003 & $2.475 \times 10^{-8}$ & 0.0366 \\
 CMB-S4 + DESI + SKA1 IM2 & 0.00003 & 0.00015 & 0.00460 & 0.00144 & 0.071 & 0.00256 & $3.175 \times 10^{-8}$ & 0.0125 \\
 \hline\hline

\end{tabular}
\caption{The expected 1$\sigma$ uncertainties on the cosmological parameters for various future CMB experiments LiteBIRD, COrE-M5, CMB-S4, PICO, Low-l LiteBIRD with High-l CMB-S4, Low-l COrE-M5 with High-l CMB-S4, CMB-S4 + DESI, CMB-S4 + Euclid (CS + GC) + SKA1 Intensity Mapping Band 2, CMB-S4 + Euclid (CS + GC), CMB-S4 + SKA1 (CS + GC), CMB-S4 + SKA2 (CS + GC) and CMB-S4 + DESI + SKA1 Intensity Mapping Band 2. For the above analysis we have considered the Conservative error approach for CS and Realistic error approach for GC of Euclid, SKA1 and SKA2 future experiments.}
\label{error2}
\end{table*}

\section{Fisher forecast on future CMB missions: A brief investigation}
\label{cmb-mission}

Till now we have taken the latest constraints from Planck18 as the sole CMB missions. 
Having convinced ourselves on the prospects of future 21-cm and galaxy surveys, let us now very briefly examine if adding future CMB missions help in improvements in predicted error.
The upcoming CMB missions  that we would like to explore  are as follows:
\begin{itemize} 
\item LiteBIRD: It is a space-based mission \citep[]{Matsumura:2013aja}, \citep[]{Suzuki:2018cuy} focused to study the primordial B-mode polarization and inflation over the entire sky with sensitivity $\delta r < 0.001$ in tensor to scalar ratio.

\item COrE-M5: COrE-M5 \citep[]{CORE:2017oje} is satellite mission with same science goal as LiteBIRD but with higher resolution.

\item CMB-S4: It is a ground based CMB experiment \citep[]{2016arXiv161002743A}, \citep[]{2017arXiv170602464A} which has greater sensitivity and resolution compared to LiteBIRD and COrE-M5. 
\item PICO: It is a satellite based mission \citep[]{Sutin:2018onu}, \citep[]{Young:2018aby} of NASA with improved sensitivity compared to LiteBIRD.
\end{itemize}

\begin{figure}
	\includegraphics[width=\columnwidth]{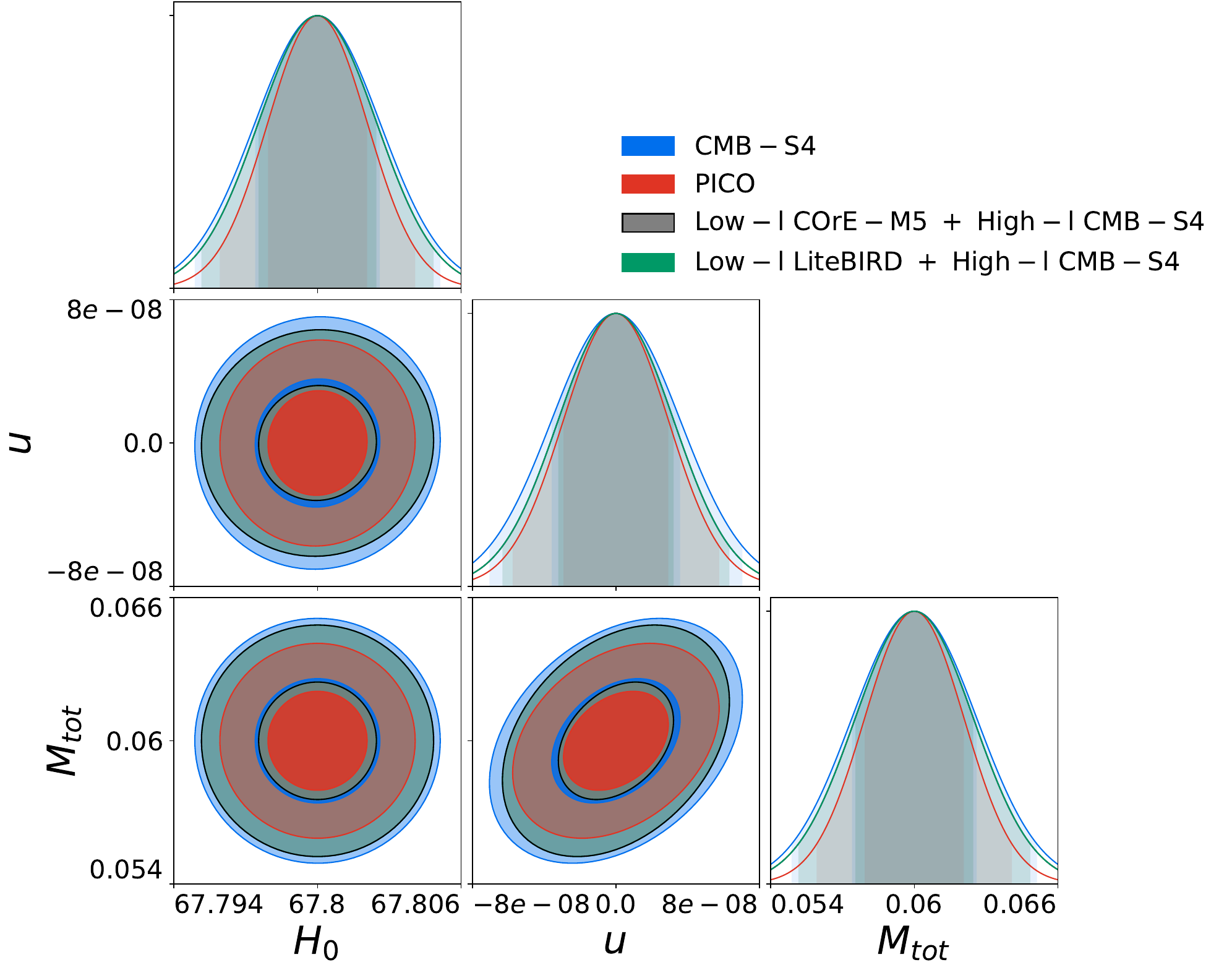}
	\caption{ From Fisher matrix analysis the above figure shows marginalized 1$\sigma$ and 2$\sigma$ contours and one-dimensional posteriors for the cosmological parameters using future CMB experiments CMB-S4, PICO, Low-l CMB-S4 $\&$ High-l COrE-M5 and Low-l LiteBIRD $\&$ High-l CMB-S4.}

\label{CMB_Full}
\end{figure}

\begin{figure}
	\includegraphics[width=\columnwidth]{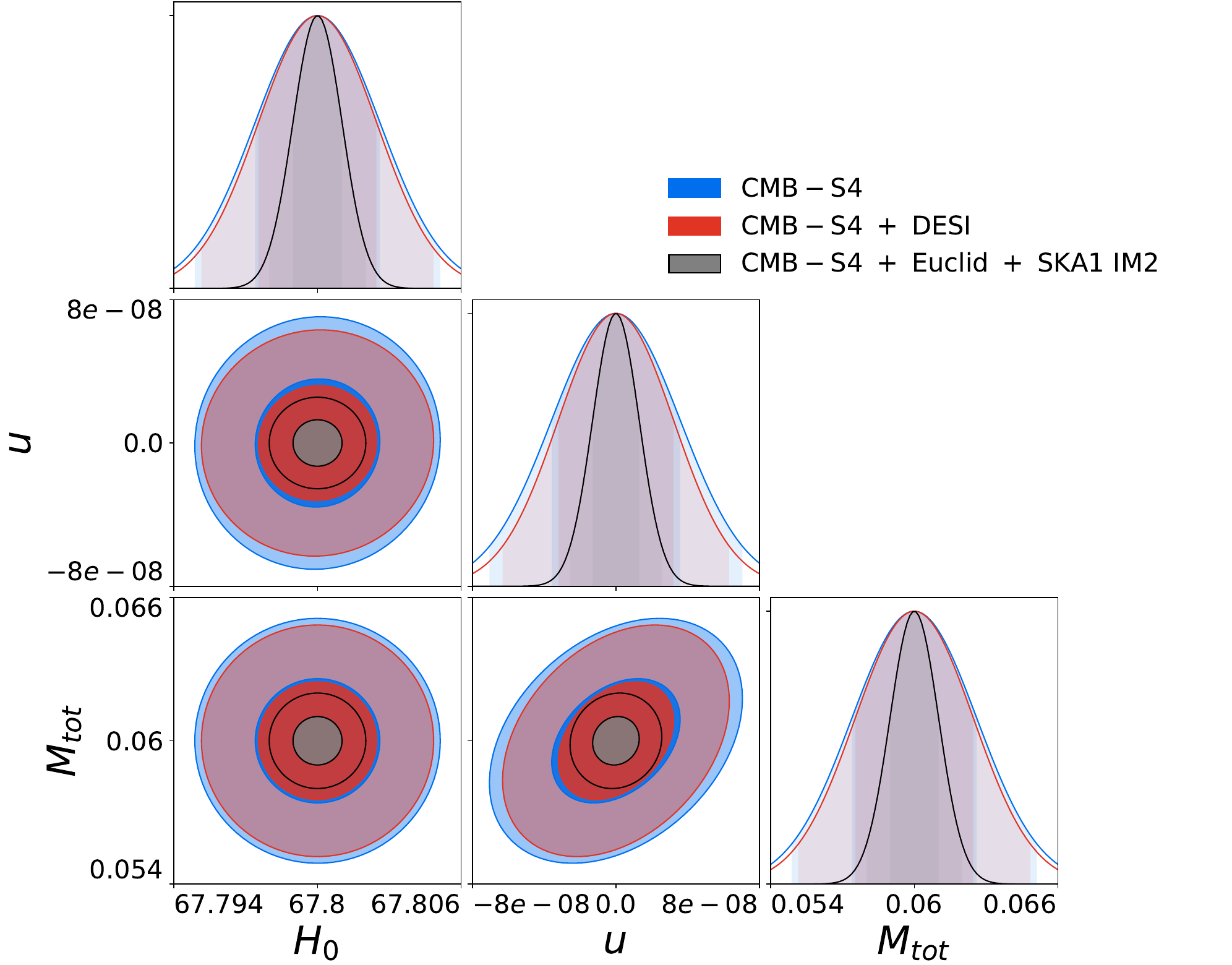}
	\caption{ From Fisher matrix analysis the above figure shows marginalized 1$\sigma$ and 2$\sigma$ contours and one-dimensional posteriors for the cosmological parameters using future CMB experiments CMB-S4, CMB-S4 + DESI and CMB-S4 + Euclid (CS + GC) + SKA1 Intensity Mapping Band 2.}
\label{CMB-desi-ska1IM}
\end{figure}

\begin{figure}
	\includegraphics[width=\columnwidth]{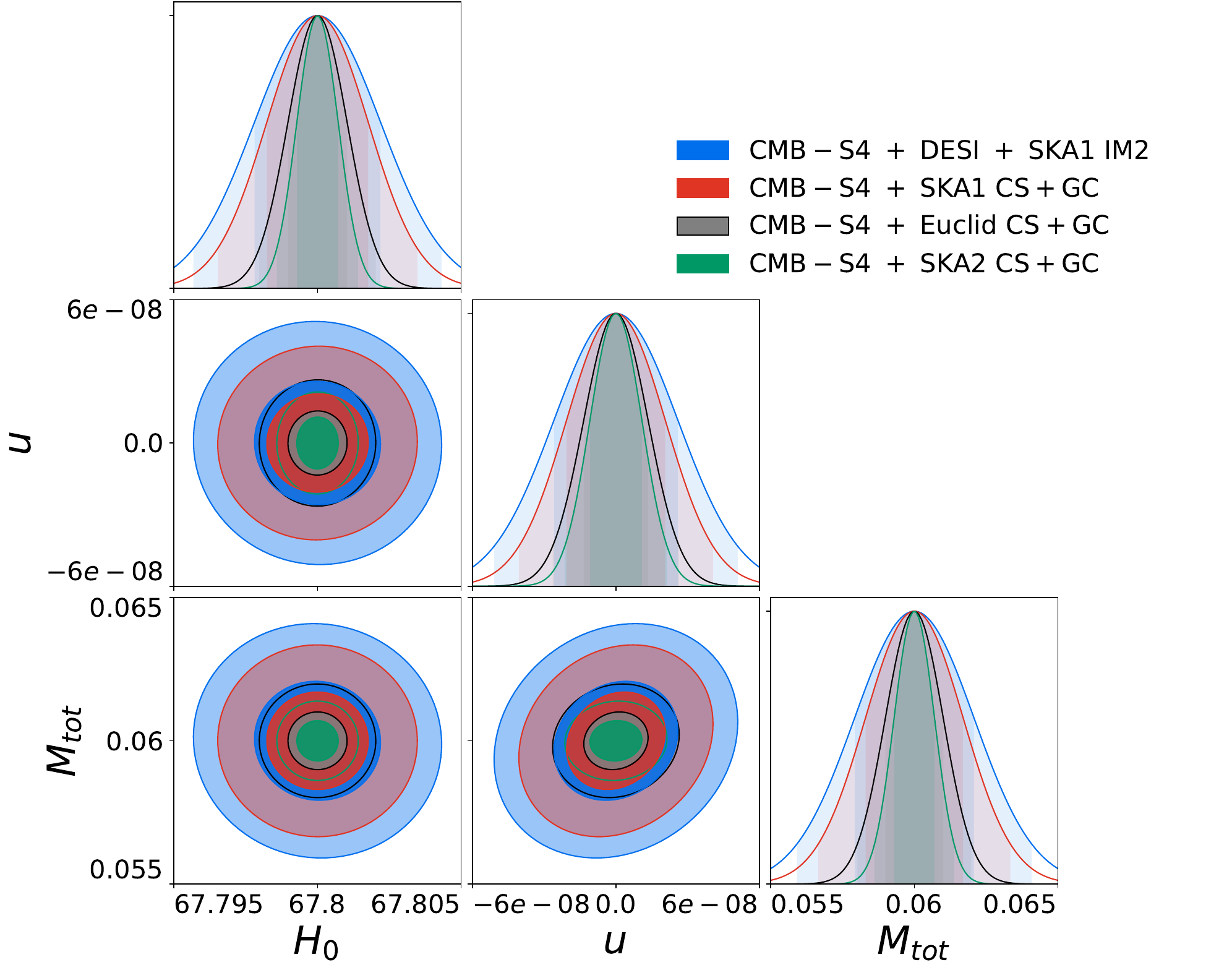}
	\caption{From Fisher matrix analysis the above figure shows marginalized 1$\sigma$ and 2$\sigma$ contours and one-dimensional posteriors for the cosmological parameters using future CMB experiments CMB-S4 + DESI + SKA1 Intensity Mapping Band 2, CMB-S4 + SKA1 (CS + GC), CMB-S4 + Euclid (CS + GC) and CMB-S4 + SKA2 (CS + GC).}
\label{CMBs4-desi-euclid-ska1IM}
\end{figure}

\begin{figure}
	\includegraphics[width=\columnwidth]{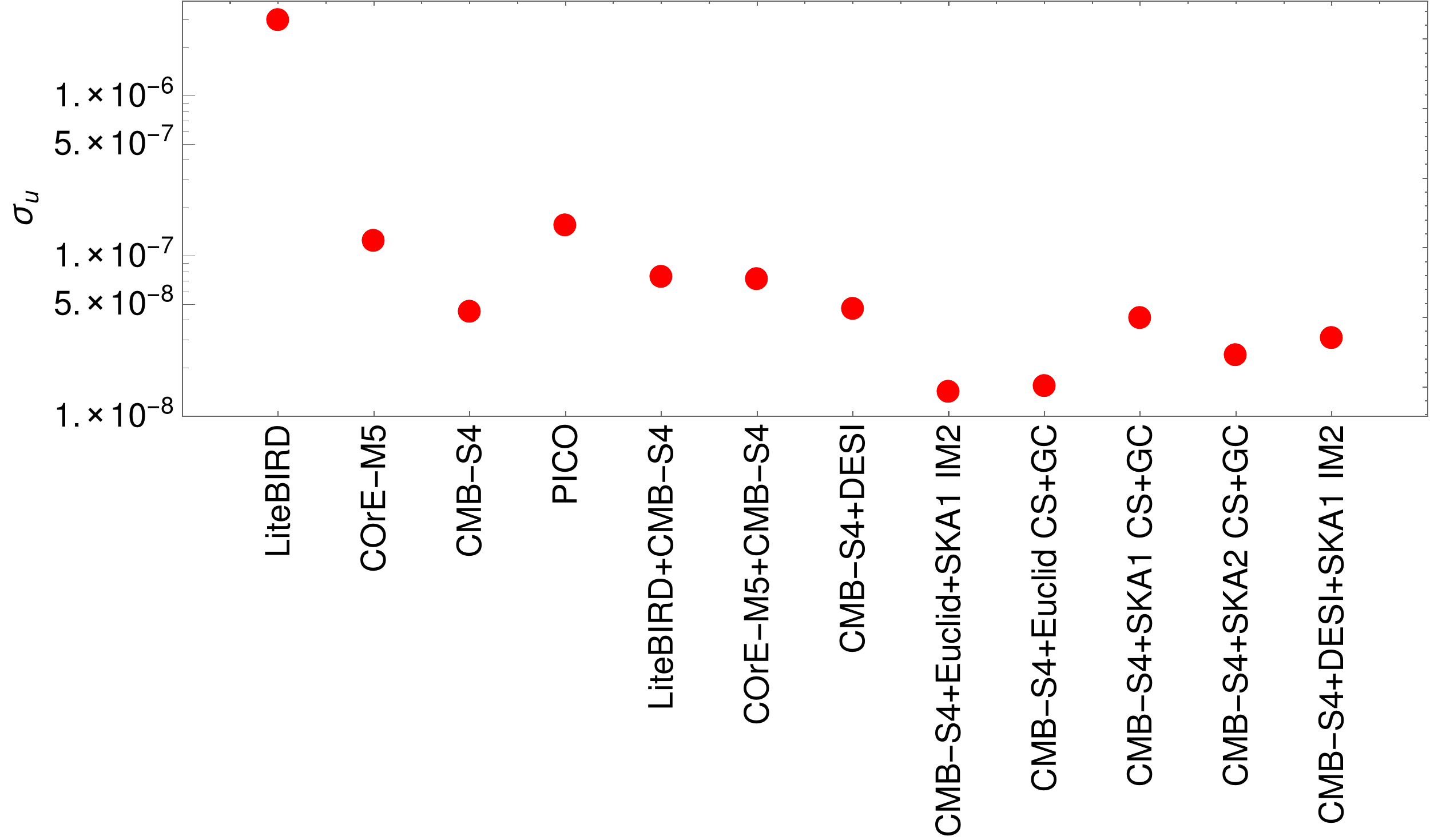}
    \caption{The above figure depicts 1$\sigma$ uncertainties on the DM-massive $\nu$ interaction parameter `$u$' for various upcoming CMB and LSS experiments.}

\label{sigmas_u_CMB_all}
\end{figure}

\begin{figure}
	\includegraphics[width=\columnwidth]{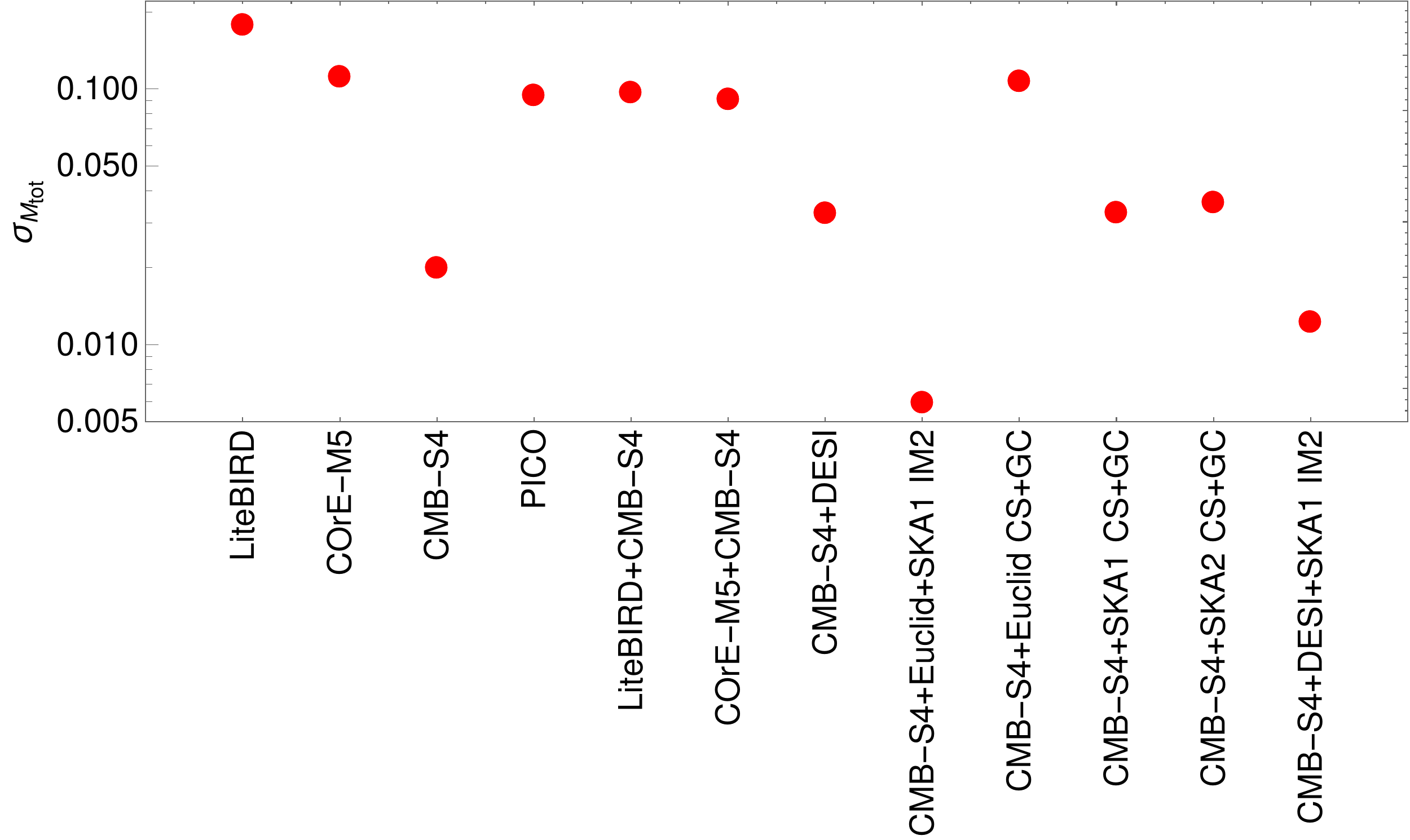}
	\caption{The above figure depicts 1$\sigma$ uncertainties on the mass of $\nu$'s for various upcoming CMB and LSS experiments.}
\label{sigmas_mtot_cmb}
\end{figure}

We have done Fisher forecast analysis of our model parameters using the future CMB missions and different combinations of  CMB and LSS experiments. The $1\sigma$ error on the model parameters has been depicted in Table \ref{error2} and error ellipse has been plotted in Figure \ref{CMB_Full}, \ref{CMB-desi-ska1IM} $\&$ \ref{CMBs4-desi-euclid-ska1IM}. The 1$\sigma$ errors on `$u$' $\&$ `$M_{\rm tot}$' from a range of future CMB experiments are illustrated in Figure \ref{sigmas_u_CMB_all} $\&$ \ref{sigmas_mtot_cmb} providing a graphical description of the results. From all the future LSS and CMB experiments that we have considered in our analysis, a combination of CMB-S4, Euclid $\&$ SKA1 IM2 provides the strongest $1\sigma$ upper bound on the interaction parameter $u\leq1.465 \times 10^{-8}$ and $\sigma(M_{\rm tot}) \sim 0.006$ eV which is of the same order of some other combination of missions. Overall, CMB experiments can put constraints on `$u$' up to an order of $\sim 10^{-8}$. The model parameters are subject to less stringent constraints when considering alternative combinations of CMB and LSS experiments. The chances of successfully detecting the interaction between DM and massive neutrinos are notably high, thanks to the forthcoming SKA, Euclid, and CMB-S4 projects taken together.

\section{Summary} \label{Summary}
In the preparatory phase of future survey Square Kilometre Array and Euclid which has launched recently, it is crucial to generate reliable and precise forecasts regarding the sensitivity to cosmological parameters. In this work, we have investigated thoroughly the effect of DM-massive neutrino interaction in post-reionization epoch using forthcoming missions SKA $\&$ Euclid and their possible combinations. The interaction between DM and neutrinos has been widely investigated in the literature and previous studies have placed constraints on the parameter space based on CMB observations and reionization physics. As a first attempt, we have conducted this analysis during the post-reionization epoch to constrain the parameter space based on forthcoming missions. In order to analyze the scenario in more concrete language, we have taken into account two approaches namely, the conservative and the realistic error estimation approaches based on the non-linear scale cut-off. Our findings highlight the significant impact of errors on the constraints of the model parameters. Notably, when conducting a joint analysis and considering combinations of two errors, we observe a substantial tightening of the bounds on the cosmological parameters. We utilize a dual approach consisting of Fisher matrix forecast analysis and MCMC simulation to thoroughly investigate the constraints and signatures of the scenario on forthcoming post-reionization, cosmic shear and galaxy surveys. Further, we have done a very brief investigation on the prospects of next generation CMB missions and combination of CMB and LSS missions in this context. We have observed that, the DM-massive neutrino interaction parameter can be constrained significantly, stronger to about 3-4 orders of magnitude than the current constraint \citep[]{Mosbech:2020ahp} to be precise, once the data is available. This is true for both upcoming LSS and CMB missions.

Our analysis primarily focuses on two key parameters: the strength of the interaction ($u$) between dark matter $\&$ massive neutrinos and total neutrino mass ($M_{\rm tot}$). These parameters are examined alongside the usual set of 6 cosmological parameters. The results of our investigation indicate that the most stringent constraints on the dark matter neutrino interaction parameter is $u\leq10^{-8}$ and the sum of neutrino masses is $\sigma(M_{\rm tot})\sim 0.006$ eV, which are achieved by combining SKA2 cosmic shear $\&$ galaxy clustering and CMB-S4 with Euclid, in conjunction with SKA1 IM2. The limits imposed on these model parameters are exceptionally robust, exceeding the previous constraints by a factor of four orders of magnitude \citep[]{Mosbech:2020ahp}. Previous investigations indicate that $\sigma(M_{\rm tot})$ is around 0.1 $\pm$ 0.0022 eV for SKA1 MID and BOSS-like Lyman-$\alpha$ forest surveys \citep[]{Pal:2016icc}, and 0.012 eV, 0.015 eV, and 0.025 eV for Planck18 + SKA2, Planck18 + Euclid + SKA1 Intensity Mapping Band 2 and Planck + SKA1 experiments \citep[]{Sprenger:2018tdb}. In our present analysis, we have found $\sigma(M_{\rm tot}) \sim 0.006$ eV in presence of dark matter neutrino interaction. Therefore, the upcoming LSS (and CMB) missions hold the potential to provide more conclusive evidence regarding the interaction between these two cosmic species and massive neutrinos. It can offer valuable insights into the nature and strength of this interaction, complementing our knowledge derived from particle physics theories and experiments. Furthermore, these findings will enhance our understanding of these cosmic entities in addition to the existing cosmological observations, such as the CMB.

\section*{Acknowledgements}

Authors gratefully acknowledge the use of publicly available code \texttt{CLASS} 
\citep[]{Blas_2011}, \texttt{MontePython} v3.4 \citep[]{Audren:2012wb}, \citep[]{Brinckmann:2018cvx} and thank the computational facilities of Indian Statistical Institute, Kolkata.  AD and AP would like to thank Debabrata Chandra for useful discussions. AD thanks ISI Kolkata for financial support through Senior Research Fellowship. AP thanks IACS, Kolkata for financial support through Research Associateship. SP thanks the Department of Science and Technology, Govt. of India for partial support through Grant No. NMICPS/006/MD/2020-21.

\section*{Data Availability}

The simulated data underlying this work will be shared upon reasonable request to the corresponding author. 



\bibliographystyle{mnras}
\bibliography{references}




\bsp	
\label{lastpage}
\end{document}